\documentclass[manuscript]{acmart}
\usepackage{subcaption}

\AtBeginDocument{%
  }

\setcopyright{acmlicensed}
\copyrightyear{2024}

\acmISBN{978-1-4503-XXXX-X/18/06}

\usepackage{rotating}
\usepackage{longtable}
\usepackage{booktabs}
\usepackage{tabularx}
\usepackage{adjustbox}
\usepackage{lscape}
\usepackage[labelfont=bf]{caption} 
\usepackage{graphicx}
\usepackage{float}
\usepackage{geometry}
\usepackage{array}
\usepackage{ragged2e}
\usepackage{lipsum}
\usepackage[utf8]{inputenc}
\usepackage[T1]{fontenc}
\usepackage[normalem]{ulem}
\useunder{\uline}{\ul}{}
\usepackage{pifont} 
\usepackage{multirow}
\usepackage{makecell} 
\usepackage{needspace}

\begin{document}

\title{Chaos Engineering: A Multi-Vocal Literature Review}

\author{Joshua Owotogbe}
\email{j.s.owotogbe@tilburguniversity.edu}
\orcid{0000-0001-9865-9310}
\author{Indika~Kumara}
\email{i.p.k.weerasinghadewage@tilburguniversity.edu}
\orcid{0000-0003-4355-0494}
\author{Willem-Jan~van~den~Heuvel}
\email{w.j.a.m.vdnHeuvel@tilburguniversity.edu}
\orcid{0000-0003-2929-413X}
\affiliation{%
  \institution{Jheronimus Academy of Data Science}
  \city{'s-Hertogenbosch}
  \state{North Brabant}
  \country{Netherlands}
}
\affiliation{%
  \institution{Tilburg University}
  \city{Tilburg}
  \state{North Brabant}
  \country{Netherlands}
}

\author{Damian~Andrew~Tamburri}
\email{datamburri@unisannio.it}
\orcid{0000-0003-1230-8961}
\affiliation{%
  \institution{Jheronimus Academy of Data Science}
  \city{'s-Hertogenbosch}
  \state{North Brabant}
  \country{Netherlands}
}
\affiliation{%
  \institution{University of Sannio}  
  \city{Benevento}
  \state{Benevento}
  \country{Italy}
}

\renewcommand{\shortauthors}{Owotogbe et al.}

\begin{abstract}

Organizations, particularly medium and large enterprises, typically rely heavily on complex, distributed systems to deliver critical services and products. However, the growing complexity of these systems poses challenges in ensuring service availability, performance, and reliability. Traditional resilience testing methods often fail to capture the intricate interactions and failure modes of modern systems. Chaos Engineering addresses these challenges by proactively testing how systems in production behave under turbulent conditions, allowing developers to uncover and resolve potential issues before they escalate into outages. Though chaos engineering has received growing attention from researchers and practitioners alike, we observed a lack of reviews that synthesize insights from both academic and grey literature. Hence, we conducted a Multivocal Literature Review (MLR) on chaos engineering to address this research gap by systematically analyzing 96 academic and grey literature sources published between January 2016 and April 2024. We first used the chosen sources to derive a unified definition of chaos engineering and to identify key functionalities, components, and adoption drivers. We also developed a taxonomy for chaos engineering platforms and compared the relevant tools using it. Finally, we analyzed the current state of chaos engineering research and identified several open research issues. 

\end{abstract}




\maketitle

\section{Introduction}

Cloud-based applications are being adopted across various domains, including healthcare, finance, construction, education, and defense~\cite{alabbadi2011cloud,dang2019survey,BELLO2021103441}. While cloud computing promises to offer benefits such as cost savings, flexibility, and faster time to market~\cite{BUYYA2009599}, the growing complexity of cloud infrastructure poses a significant challenge to maintaining dependable systems, leading to an increased risk of technical failures and costly downtime~\cite{song2022influence,li2011trusted,wang2015environment, al2024exploring}.
The consequences of these failures can be severe, as seen in significant financial losses incurred for companies, including the 2007-2013 outages of 38 cloud service providers, resulting in \$669 million in losses, and Facebook's six-hour outage in 2021, which resulted in an estimated \$160 million loss for the global economy ~\cite{Gill2022}. The 2017 ITIC (Information Technology Intelligence Consulting) survey also reported that 98\% of organizations experienced downtime with financial losses exceeding \$100,000 per hour~\cite{Gremlin23}.

An effective approach to overcoming the challenges of ensuring reliable and resilient cloud-based systems is chaos engineering, which proactively tests system resilience by introducing controlled failures into production environments~\cite{Bhaskar22, Fawcett2020, Palani23}. This approach was pioneered by Netflix in 2008 to fortify its video streaming infrastructure. Since then, chaos engineering has grown and extended its influence, evolving from its original focus on cloud services and microservice architecture to broader applications across industries~\cite{Bairyev2023}. Today, this approach is widely employed in critical sectors, such as fintech and healthcare, with major technology companies like LinkedIn, Facebook, Google, Microsoft, Amazon, and Slack adopting it to enhance system resilience, improve customer experience, and drive business growth. ~\cite{Gremlin23, Patel22, Roa2022, Fawcett2020, al2024exploring, Bremmers21,Bairyev2023}.



We observed that many researchers and practitioners have explored chaos engineering in various contexts, discussing its implementations and challenges through research papers and grey literature such as online blogs, technical papers, and forums. However, despite its increasing adoption, much of the existing knowledge remains dispersed across the literature. To the authors' best understanding, no secondary research has been published to synthesize this scattered information, such as comprehensive reviews or survey papers. Hence, to address this research gap, in this paper, we present a Multivocal Literature Review (MLR)~\cite{garousi2019guidelines} on chaos engineering, synthesizing a broad spectrum of perspectives from academic and grey literature sources. Unlike traditional Systematic Literature Reviews (SLRs), which often overlook grey literature, an MLR includes insights from peer-reviewed academic studies alongside industry reports, technical blogs, and white papers~\cite{garousi2019guidelines,garousi2016need}. In particular, as chaos engineering emerged from the industry and most tools are from the industry, the grey literature is a highly relevant and valuable source of information. In this MLR, we selected and analyzed 96 sources, encompassing both academic and grey literature. 
We selected 2016 as the starting point, when chaos engineering principles were first formally articulated by Basiri et al~\cite{basiri2016chaos}.
Our findings provide a detailed overview of chaos engineering, including its core principles, quality requirements, and foundational components defined by researchers and industry experts. We also examine the motivations behind adopting chaos engineering, the specific resilience challenges it addresses, and the practices and evaluation methods that support its effective implementation.
Additionally, we introduce a taxonomy of chaos engineering platforms and their capabilities, designed to support informed decision-making around tool adoption and deployment strategies.
By analyzing the benefits and shortcomings of current approaches in the chaos engineering domain, this MLR contributes to a thorough understanding of chaos engineering and provides guidelines for future studies to advance this critical field.

The remainder of the paper is organized as follows. Section~\ref{sec:Background} outlines the background and motivation for this review, including a discussion of related surveys, the distinction between chaos engineering and traditional testing, and the need to extend the existing body of knowledge on chaos engineering. Section~\ref {sec:Method} presents the research methodology used in this study, while Section~\ref{sec:Definition} provides a comprehensive definition of chaos engineering, along with its functionalities, quality requirements, and core components. Section~\ref{sec:solved_challenges} focuses on the motivations for chaos engineering and the specific challenges it addresses. Section~\ref{sec:Framework} introduces a taxonomy of chaos engineering, covering tools, practices, and evaluation methods within the field. Section~\ref{sec:Current_state} examines the current state of chaos engineering in research and industry. Section ~\ref{sec:Discussion} discusses the implications of our findings, highlighting key insights, identifying open issues, and addressing potential threats to validity. Finally, Section ~\ref{sec:Conclusion} concludes the paper.

\section{Background and Motivation}
\label{sec:Background}

This section outlines the motivations for conducting this review by examining the limitations of existing literature (Section\ref{sec:relatedsurvys}), explaining how chaos engineering diverges from traditional testing methods (Section \ref{sec:Chaosdistinct}), and establishing the novelty and timeliness of this multivocal literature review (Section \ref{sec:Knowledgebody}).

\subsection{Related Work and Survey Gap}
\label{sec:relatedsurvys}

In this section, we consider key surveys on resilience, robustness, and fault injection techniques, exploring their relationship to the proactive focus of chaos engineering. For instance, Natella et al.~\cite{natella2016assessing} surveyed software fault injection techniques, categorizing faults by origin such as software, hardware, and environment, and by injection approach such as code changes or fault effect injection including state and interface errors. However, their focus is on pre-deployment environments (e.g., simulators, testbeds), whereas chaos engineering involves live, iterative experiments in production-like systems. Similarly, Colman-Meixner et al.~\cite{colman2016survey} focus on cloud resilience through redundancy and failover, but explore reactive recovery rather than proactive experimentation, as in chaos engineering.
Laranjeiro et al.~\cite{laranjeiro2021systematic} consider static robustness techniques such as fuzzing and API testing, which lack real-time, adaptive feedback loops. Ruholamini et al. ~\cite{rouholaminiproactive} surveyed self-healing in cloud-native systems, emphasizing failure detection and recovery rather than proactive failure exploration.

More recently, Arsecularatne and Wickramarachchi~\cite{arsecularatne2023adoptability} conducted a systematic review and qualitative study on the adoptability of chaos engineering in DevOps. Their work highlights organizational challenges and potential benefits but does not provide a structured analysis of tools, architectures, or evaluation metrics. Similarly, Mhatre et al.~\cite{mhatrerole} examined chaos engineering within DevSecOps and SRE (Site Reliability Engineering) contexts, emphasizing conceptual alignment and use cases, but without offering a comprehensive taxonomy or resilience assessment frameworks. In summary, existing reviews examine fault injection or discuss chaos engineering within DevOps contexts. However, they offer little or no clear insight into its unified definition, functionalities, architectural components, quality attributes, comprehensive tool taxonomy, or the technical and organizational challenges of implementing it in production-like environments. This review addresses these gaps by presenting the first multivocal literature review (MLR) on chaos engineering, integrating both academic and industry sources to inform future research and practical adoption.

\subsection{Chaos Engineering vs. Traditional Software Testing Techniques}
\label{sec:Chaosdistinct}

Traditional software testing techniques, including unit, integration, stress, and robustness testing, primarily focus on validating correctness and performance under expected conditions in pre-deployment environments~\cite{basiri2019automating,shekhar2024chaos}. These approaches typically operate in isolated, controlled settings and verify that system components behave correctly with known inputs~\cite{ji2023perfce,shekhar2024chaos}. However, in modern distributed, cloud-native systems, such methods often fail to expose cascading or emergent failures caused by asynchronous interactions and dynamic runtime behaviors~\cite{migirditch2022chaos,ji2023perfce,fogli2023chaos}.
Chaos engineering addresses these limitations by introducing controlled faults into production or production-like systems to explore system resilience~\cite{zhang2019chaos,torkura2020cloudstrike,fogli2023chaos}. Rather than validating functionality using predefined test scripts, chaos experiments simulate failures, such as service crashes, latency spikes, or regional outages,to evaluate system behavior and recovery under adverse runtime conditions~\cite{zhang2019chaos,migirditch2022chaos,shekhar2024chaos}. This methodology helps surface hidden vulnerabilities that traditional test environments often miss~\cite{ji2023perfce,parsons2021chaos}. Importantly, chaos engineering is not a replacement for conventional testing, but a complement. While traditional methods confirm system correctness under expected conditions, chaos engineering validates system behavior under known faults and real-world failure scenarios ~\cite{basiri2019automating,fogli2023chaos}. This paradigm shift—from pre-deployment pass/fail validation to post-deployment resilience assessment—is what makes chaos engineering essential for today's complex, distributed infrastructure~\cite{zhang2019chaos,ji2023perfce,fogli2023chaos}.
To clarify these distinctions, Table~\ref{tab:traditional_vs_chaos} highlights key differences between traditional testing and chaos engineering. The comparison focuses on three main areas: the testing environment (pre-production vs. production-like systems), the primary objective (functional verification vs. uncovering hidden system weaknesses), and the types of failures addressed (expected vs complex real-world faults).


\begin{table}[htbp]
\caption{Comparison Between Traditional Software Testing Techniques and Chaos Engineering}
\label{tab:traditional_vs_chaos}
\small
\begin{tabular}{|p{0.12\textwidth}|p{0.30\textwidth}|p{0.30\textwidth}|p{0.16\textwidth}|}
\hline
\textbf{Category} & \textbf{Traditional Testing} & \textbf{Chaos Engineering} & \textbf{Articles Referenced} \\
\hline
\textbf{Testing Environment} & Pre-production or simulated systems. & Production or production-like systems. & \cite{Bairyev2023,zhang2019chaos,simonsson2021observability,zhang2021maximizing,Gill21,Bocetta19,Bhaskar22,Krivas2020,Wilms2018}
 \\
\hline
\textbf{Testing Objective} & Verifies specific functions or components against pre-defined requirements and standards. & Identifies obscure system weaknesses that may only become apparent during turbulent failure conditions. & \cite{torkura2020cloudstrike,Gremlin23,Bairyev2023} \\
\hline
\textbf{Types of Failures} & Focuses on expected and known usage scenarios and errors to verify how the system behaves under normal or predictable conditions. & 
Simulates real-world failures   (e.g., network outages, high load, hardware issues) to see how the system reacts to these changes.
& \cite{torkura2020cloudstrike,pierce2021chaos,ji2023perfce,konstantinou2021chaos,Dua24,Miles19,Bairyev2023}
 \\
\hline
\end{tabular}
\end{table}

\subsection{Research Contribution and Novelty}
\label{sec:Knowledgebody}

Despite the growing popularity of chaos engineering~\cite{monroeinvestigate}, there remains a lack of unified academic frameworks that synthesize its principles, classify its tool landscape, or integrate its practices across both research and industry domains. To the best of our knowledge, this paper presents the first Multivocal Literature Review (MLR) on chaos engineering, integrating both academic studies and grey literature.
Our contributions are fourfold. First, we synthesize and unify definitions, core principles, functionalities, and architectural components of chaos engineering from academic and industry perspectives \textbf{(RQ1)}. Second, we identify the key technical and socio-technical challenges driving its adoption \textbf{(RQ2)}. Third, we propose a taxonomy of chaos engineering tools, classify evaluation methods, and highlight adoption practices \textbf{(RQ3)}. Fourth, we analyze publication trends, venue types, and key contributors to assess the current state of research and practice \textbf{(RQ4)}. Together, these contributions provide a foundational reference on chaos engineering that bridges research and practice, supports rigorous evaluation of chaos engineering tools,  enables their systematic adoption in industry, and offers directions for future research.

\section{Research Methodology}
\label{sec:Method}
\begin{figure}[htbp]
  \centering
  \vspace{-10pt}\includegraphics[width=0.70\textwidth]{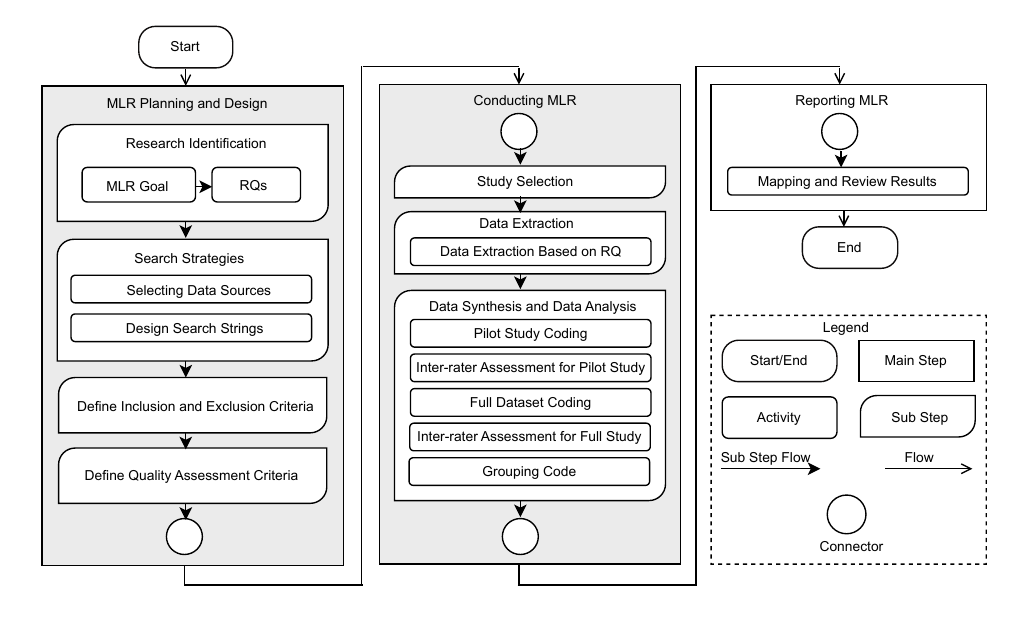}
  \caption{Overview of the MLR Process~\cite{islam2019multi}. }
  \label{fig:CE_MLR_PROCESS}
  \Description{Multivocal Literature Review Process.}
\end{figure}

We employed the guidelines proposed by Garousi et al.~\cite{garousi2019guidelines} to systematically collect and analyze multivocal literature on chaos engineering.  We also drew methodological inspiration from other multivocal literature reviews, such as those by Buck et al.~\cite{buck2021never} and Islam et al.~\cite{islam2019multi}.
Figure~\ref{fig:CE_MLR_PROCESS} illustrates our MLR methodology, comprising several steps divided into three phases: planning and designing the review protocol, executing the review, and documenting the findings. The remainder of this section provides a detailed discussion of each step. 

\subsection{Research Identification}
We employed a search strategy to identify appropriate literature grounded in a series of research questions in Table~\ref {tab:research_questions}. The first three questions cover the various aspects of chaos engineering, including its definitions, key features, benefits, challenges, use cases, tools and techniques, and best and bad practices. The final question provides an overview of the present landscape of chaos engineering literature.

\begin{table}[htbp]
  \caption{Research Questions and Motivation}
  \label{tab:research_questions}
   \small
  \begin{tabular}{|p{0.55\textwidth}|p{0.35\textwidth}|}
    \hline
    \textbf{Research Questions} & \textbf{Motivation} \\
    \hline
    \textbf{RQ1}. How is chaos engineering defined in the literature? 
    
    \textit{RQ1.1.} Which core activities and primary functionalities are most commonly associated with chaos engineering?
    
    \textit{RQ1.2.} What are the core components of a chaos engineering platform, and how do they support its core functions?
    
    \textit{RQ1.3.} What quality requirements are essential for an effective chaos engineering platform, and how are they quantified?
    & 
 This question clarifies the definition of chaos engineering, exploring its core activities, key components, and quality requirements. These elements build a foundation for understanding, implementing, and evaluating chaos engineering solutions.
 \\
    \hline
    \textbf{RQ2}. What challenges does chaos engineering intend to solve? 

    &
    This question helps identify key challenges, guiding organizations to areas where chaos engineering can best enhance resilience and reliability. \\
    \hline
    \textbf{RQ3}. What kinds of solutions have been proposed for chaos engineering?
    
      \textit{RQ3.1.} What tools and techniques are utilized, proposed, designed, and implemented by researchers and practitioners?
    
      \textit{RQ3.2.} What best and bad practices are documented in adopting chaos engineering?
    
      \textit{RQ3.3.} How are the chaos engineering approaches evaluated?&
    This question examines the tools and techniques in chaos engineering, highlighting key practices, common pitfalls, and methods for evaluating and enhancing solutions across various contexts. \\
    \hline
    \textbf{RQ4}. What is the state of chaos engineering in research and industry?

     \textit{RQ4.1.} What is the extent and diversity of research conducted on chaos engineering, and what trends can be observed over time?

      \textit{RQ4.2.} Who are the thought leaders and key contributors in the field of chaos engineering, and what intersections of thought do they represent?

     &
    Exploring this question allows us to assess the growth and direction of chaos engineering, highlighting current trends, influential contributors, and areas that need further exploration, ensuring continued advancement in the field. \\
    \hline
  \end{tabular}
\end{table}

\subsection{Search Strategy}
This section outlines the search strategy, detailing the data sources, queries, and the complete search process. 

\subsubsection{Sources of data}
We conducted automated searches in four major digital libraries—Scopus, IEEE Xplore, ACM Digital Library, and Web of Science—using both peer-reviewed scholarly articles and grey literature as sources. These libraries serve as primary sources for software engineering studies~\cite{chen2010towards}. 
Following the formal articulation of chaos engineering principles by Basiri et al. in 2016~\cite{basiri2016chaos}, we limited our search to publications from January 2016 to April 2024. We performed multi-keyword searches over titles, abstracts, and keywords within this range. Additionally, Google Scholar was used to pinpoint any relevant literature that was not included in the previous searches. We used Google Search for grey literature, similar to other multi-vocal literature reviews (MLRs)~\cite{garousi2016and,garousi2017software,islam2019multi}.

\subsubsection{Search Terms}

Following the SLR guidelines~\cite{kitchenham2007guidelines}, we constructed search terms based on the research questions. The first two authors tested the search strings to assess their applicability for addressing the defined research questions. Multiple pilot searches were conducted to refine the search terms, ensuring the inclusion of all relevant papers—boolean operators (AND, OR) combined keywords, synonyms, and related concepts. Table~\ref{tab:search_results} shows the final search strings used across the four digital libraries. We adapted and applied the Google Search search string “(Chaos AND Engineering)” for grey literature to capture relevant non-peer-reviewed sources.

\begin{table}[!htbp]
  \caption{Search Strings and Results.}
  \label{tab:search_results}
  \centering
  \begin{tabular}{|p{0.7\textwidth}|p{0.5cm}|p{0.5cm}|p{0.5cm}|p{0.5cm}|p{0.5cm}|}
    \hline
    \textbf{Search Strings} 
    & \rotatebox{90}{\shortstack{\textbf{ACM}\\\textbf{DL}}} 
    & \rotatebox{90}{\shortstack{\textbf{IEEE}\\\textbf{Xplore}}} 
    & \rotatebox{90}{\textbf{Scopus}} 
    & \rotatebox{90}{\shortstack{\textbf{Web of}\\\textbf{Science}}} 
    & \rotatebox{90}{\textbf{Total}} \\
    \hline
    "chaos engineering" \textbf{OR} "chaos test" \textbf{OR} "chaos experiment" \textbf{OR} "chaos toolkit" \textbf{OR} "chaos platform" \textbf{OR} "chaos mesh" \textbf{OR} "chaos integration" \textbf{OR} "chaos architecture" \textbf{OR} "chaos infrastructure" \textbf{OR} "chaos injection" \textbf{OR} "chaos tool" & 108 & 81 & 142 & 57 & 388 \\
    \hline
    "chaos engineering" \textbf{AND} ("trends" \textbf{OR} "developments" \textbf{OR} "progression") & 30 & 0 & 20 & 0 & 50 \\
    \hline
    "chaos engineering" \textbf{AND} ("leader" \textbf{OR} "pioneer" \textbf{OR} "expert" \textbf{OR} "contributor") & 39 & 3 & 4 & 0 & 46 \\
    \hline
    "chaos engineering" \textbf{AND} ("phases" \textbf{OR} "stages" \textbf{OR} "steps" \textbf{OR} "pipeline") & 71 & 6 & 11 & 1 & 89 \\
    \hline
    "chaos engineering" \textbf{AND} ("challenges" \textbf{OR} "issues" \textbf{OR} "problems") & 91 & 19 & 37 & 4 & 151 \\
    \hline
    "chaos engineering" \textbf{AND} ("applications" \textbf{OR} "benefits") & 89 & 22 & 43 & 9 & 163 \\
    \hline
    "chaos engineering" \textbf{AND} ("solutions" \textbf{OR} "approaches" \textbf{OR} "methods") & 84 & 22 & 60 & 4 & 170 \\
    \hline
    "chaos engineering" \textbf{AND} ("tools" \textbf{OR} "techniques") & 92 & 30 & 42 & 10 & 174 \\
    \hline
    "chaos engineering" \textbf{AND} ("best practices" \textbf{OR} "bad practices" \textbf{OR} "implementation" \textbf{OR} "Culture") & 77 & 9 & 17 & 5 & 108 \\
    \hline
    "chaos engineering" \textbf{AND} ("evaluation" \textbf{OR} "metrics") & 85 & 25 & 28 & 4 & 142 \\
    \hline
    \textbf{Total} & 766 & 217 & 404 & 94 & 1481 \\
    \hline
  \end{tabular}
\end{table}

\subsection{Eligibility Criteria}
Following some existing MLR studies~\cite{islam2019multi,butijn2020blockchains,garousi2019guidelines}, we applied a specific set of inclusion and exclusion criteria, as detailed in Table~\ref {tab:criteria}

\begin{table}[htbp]
  \caption{Inclusion and Exclusion Criteria.}
  \label{tab:criteria}
  \small
  \begin{tabular}{|p{0.45\textwidth}|p{0.50\textwidth}|}
    \hline
    \textbf{Inclusion Criteria} & \textbf{Exclusion Criteria} \\
    \hline
    \textbf{IC1:} Articles must be in English and be readily accessible in full text. & 
    \textbf{EC1:} Short articles or brief papers (less than six pages). \\
    \hline
    \textbf{IC2:} Articles must include clear validation or evaluation methods, with supporting references. & 
    \textbf{EC2:} Articles that lack the relevant focus on chaos engineering. \\
    \hline
    \textbf{IC3:} Articles that report practices and challenges in implementing chaos engineering, including case studies, best practices, or guidelines. & 
    \textbf{EC3:} Articles that predominantly discuss physical infrastructure or purely hardware aspects without direct connection to chaos engineering. \\
    \hline
    \textbf{IC4:} Articles that provide insights into the trends and evolution of chaos engineering practice. & 
    \textbf{EC4:} Studies enhancing specific algorithms or features of a single tool without a broader chaos engineering context unless they provide insights into the field's evolution or state of tools. \\
    \hline
    \textbf{IC5:} Articles that identify key contributors and diverse perspectives within chaos engineering. & 
    \textbf{EC5:} Duplicate articles from all sources. \\
    \hline
    \textbf{IC6:} Articles must cover all identified phases of the chaos engineering pipeline, ensuring a holistic view of the field. & 
    \textbf{EC6:} Articles lacking empirical evidence or practical applications to support theoretical propositions. \\
    \hline
  \end{tabular}
\end{table}

\subsection{Study Selection}
\begin{figure}[htbp]
  \centering
  \includegraphics[width=0.80\textwidth]{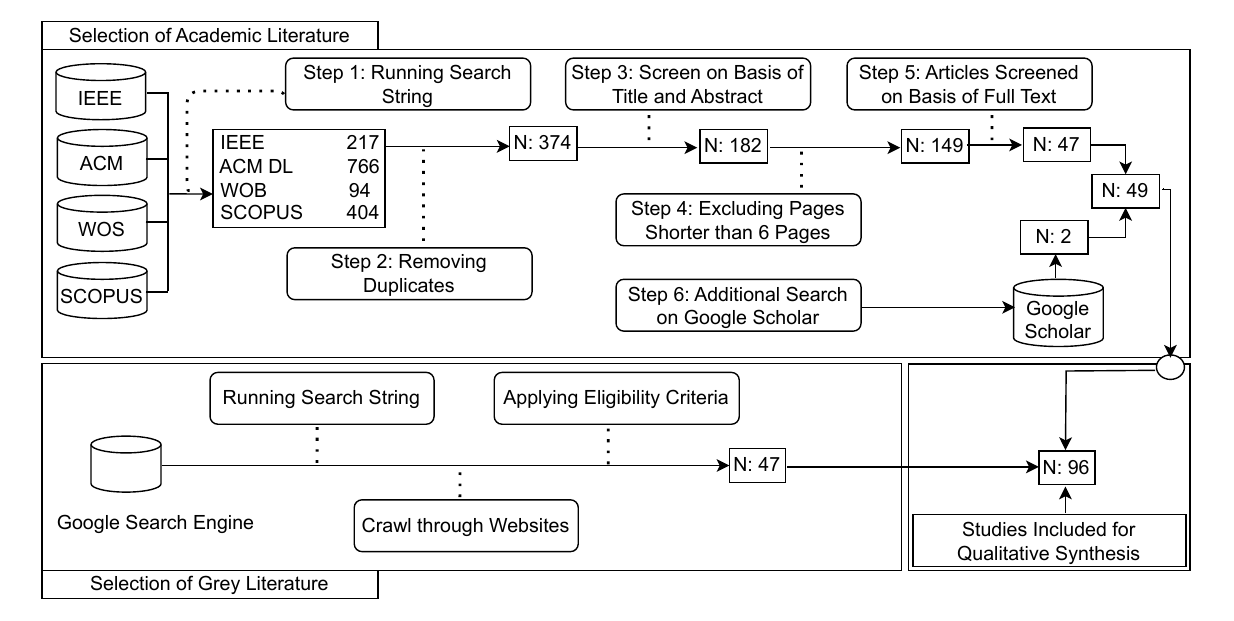}
  \caption{
  The MLR study selection process. Steps are shown sequentially for both academic and grey literature. After final screening, inter-rater reliability was assessed using Cohen's Kappa \cite{landis1977measurement}, yielding \(\kappa = 0.826\)
 for academic and \(\kappa = 0.857\)
 for grey literature, indicating strong reviewer agreement.}
  \label{fig:MULTIVOCAL_REVIEW_Final}
  \Description{MULTIVOCAL_REVIEW_Final}
\end{figure}

The libraries' advanced query modes were used to maximize search coverage. The search was conducted in April 2024, and the search results were exported from each database in BibTeX and CSV formats for reference management and data processing. Duplicates were removed before proceeding with further analysis.
Figure ~\ref{fig:MULTIVOCAL_REVIEW_Final} illustrates the selection process for academic and grey literature, showing the databases searched and the number of selected papers at each stage. Different approaches were followed to select grey and academic literature.

\subsubsection{Academic Literature Selection}

Searches in Scopus, IEEE Xplore, ACM DL, and Web of Science yielded 1,481 papers. After duplicate removal, 374 unique papers remained. Applying the inclusion and exclusion criteria, the selection was narrowed to 47 relevant papers. An additional search was conducted using Google Scholar, which identified two more articles, bringing the total to 49 papers.

\begin{table}[htbp]
\caption{Criteria for Qualitative Assessment of Grey Literature.}
\label{tab:quality_assessment}
\centering
\small
\begin{tabular}{|p{2cm}|p{10cm}|p{2cm}|}
\hline
\textbf{Category} & \textbf{Questions} & \textbf{Criteria Score} \\ \hline
\textbf{Authority} & 
(Q1). Is the organization publishing the work credible? \newline
(Q2). Are there other publications by the author in this domain of study? \newline
(Q3). How closely is the author connected to a reputable institution? & 2/3 \\ \hline
\textbf{Methodology} & 
(Q4). Is the objective of the source evident and well-stated? \newline
(Q5). Can the author be considered an expert in this field? & 2/2 \\ \hline
\textbf{Date} & 
(Q6). Is the publication date of the source within the past eight years? & 1/1 \\ \hline
\end{tabular}
\end{table}

\subsubsection{Grey Literature Selection}
We followed the approach of other multi-vocal and grey literature reviews ~\cite{garousi2016and,garousi2017software,islam2019multi,butijn2020blockchains,garousi2019guidelines} and used Google Search to gather grey literature. As described by these reviews, we examined the first ten pages of results, as this was deemed sufficient to identify the most pertinent literature since Google’s algorithm emphasizes the most relevant results on the first pages ~\cite{soldani2018pains,verdecchia2019guidelines,islam2019multi}. We continued scanning further only if necessary, stopping when no relevant new articles appeared~\cite{islam2019multi}. To avoid bias, we searched in incognito mode, following the recommendations of Rahman et al.~\cite{rahman2023defect}. 
We assessed the quality and relevance of the primary grey literature sources we obtained, as it cannot be assumed that their quality is guaranteed. The quality assessment criteria suggested by Garousi et al\cite{garousi2016and} have been used for this purpose (see Table~\ref{tab:quality_assessment}). Those criteria comprise three quality categories, ranging from the authority of the producer to the publication date (see the first column of Table~\ref{tab:quality_assessment}). These categories encompass six criteria (see the second column) that were assessed individually for each grey literature item.
The evaluation was conducted independently by the first two authors of this study, who indicated whether each item (a) met or (b) failed to meet a criterion. In cases where a criterion could not be assessed, such as when author or affiliation details were missing, it was treated as not satisfied. The final column of Table~\ref{tab:quality_assessment} specifies how many of the criteria needed to be satisfied for inclusion. For example, an item had to meet at least 2 out of 3 criteria related to the authority of the source to be considered valid. To further ensure internal agreement and consistency in applying these quality checks, all disagreements between reviewers were resolved through discussion.
As a result of this procedure, 47 grey literature sources were selected. These, together with 49 academic papers, resulted in a total of 96 sources used for data extraction and synthesis (Figure~\ref{fig:MULTIVOCAL_REVIEW_Final}). The distribution of selected sources across venues is summarized in Table~\ref{tab:distribution_literature}.

\begin{table}[htbp]
    \centering
    \caption{Selected Study for Data Synthesis and Qualitative Evaluation.}
    \label{tab:distribution_literature}
     \small
    \begin{tabular}{|p{0.39\textwidth}|p{0.39\textwidth}|p{0.12\textwidth}|}
   
        \hline
        \textbf{Academic Literature} & \multicolumn{2}{c|}{\textbf{Grey Literature}} \\ \hline
        & \textbf{Websites \& Blogs} & \textbf{White Papers} \\ \hline
        \cite{fogli2023chaos,torkura2020cloudstrike,zhang2019chaos,dedousis2023enhancing,simonsson2021observability,zhang20213milebeach,cotroneo2022thorfi,camacho2022chaos,torkura2021continuous,poltronieri2022chaos,zhang2023chaos,nikolaidis2021frisbee,frank2021interactive,frank2021scenario,pierce2021chaos,soldani2021automated,al2024exploring,vu2022mission,chen2024microfi,nikolaidis2023event,ma2023phoenix,meiklejohn2021service,frank2023verifying,sondhi2021chaos,kassab2022c2b2,jernberg2020getting,konstantinou2021chaos,pulcinelli2023conceptual,ji2023perfce,ikeuchi2020framework,chen2022big,naqvi2022evaluating,malik2023chess,bedoya2023securing,kesim2020identifying,basiri2019automating,park2019simulation,klinaku2022beauty,rivera2023using,camilli2022microservices,yu2021microrank,siwach2022evaluating,basiri2016chaos,tucker2018business,alvaro2017abstracting,sousa2018engineering,alvaro2016automating} = 49 & 
    \cite{Gremlin23,Wickramasinghe23,Gill21,Bremmers21,Patel22,Jakkaraju20,Hawkins20,Nombela23,Le22,NationalAustraliaBank20,Shah21,Krishnamurthy21,Moon22,Katirtzis22,Wachtel23,Durai22,Anwar19,Tomka24,Dua24,Bocetta19,Mooney23,Wu21,Lardo19,Miles19,Bhaskar22,AzureReadiness23,Gill2022,Fawcett2020,Manna2021,Gianchandani2022,Bairyev2023,Sbiai2023,Abdul2024,Roa2022,Kadikar2023,Green2023,Aharon2024,Krivas2020,Kostic2024,Vinisky2024,Colyer19,Hochstein2016,Butow2018,Wilms2018} = 44 & \cite{Palani23,Tivey19,Starr22} = 3 \\ \hline
    \end{tabular}
\end{table}

\subsection{Data Extraction, Synthesis, and Analysis}
\label{sec:thematic}

In alignment with the MLR process, we read and assessed the selected papers, extracting and summarizing data to address our research questions. 

\subsubsection{Data Extraction}
To systematically collect essential details, such as author names, paper titles, publication venues, and years, we used a structured data extraction form based on guidelines from Kitchenham et al.~\cite{kitchenham2004procedures} and Garousi and Felderer~\cite{garousi2017experience}. This information was organized in a spreadsheet.

\subsubsection{Synthesis and Analysis}
\begin{figure}[!htbp]
  \centering
  \vspace{-10pt}
  \includegraphics[width=0.90\textwidth]{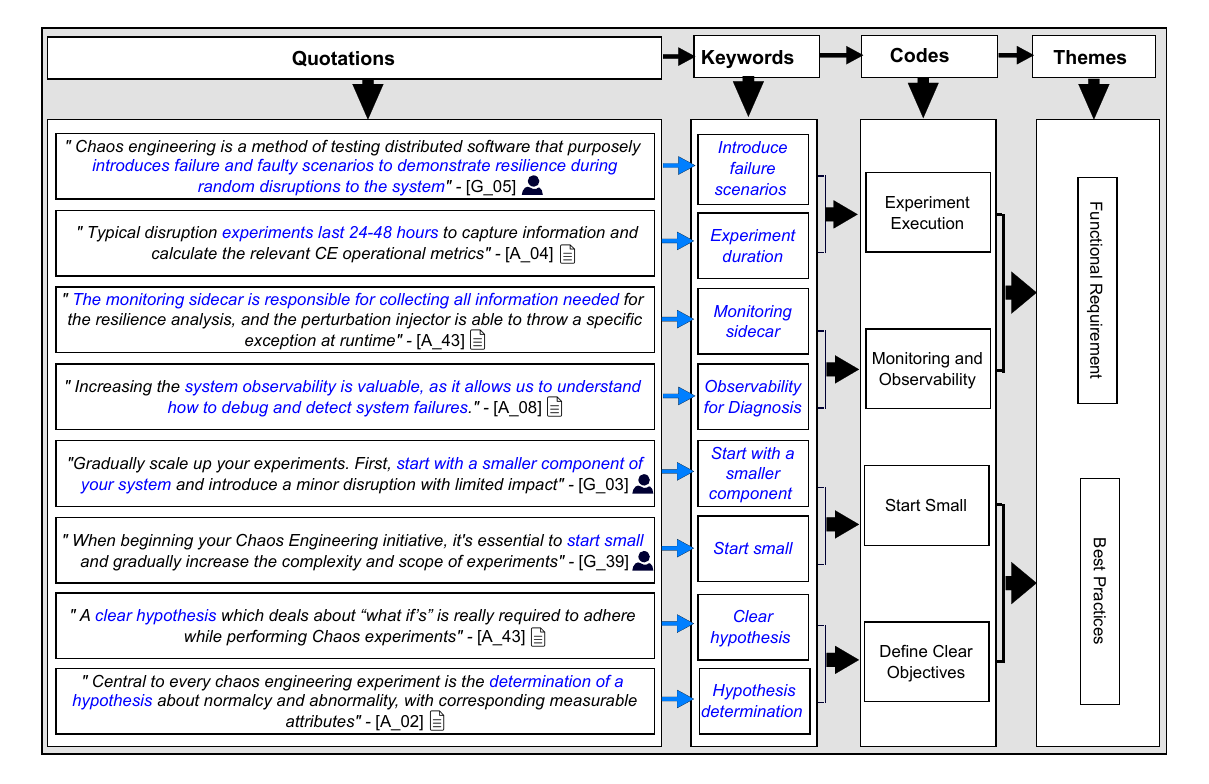}
  \vspace{-10pt}
    \caption{ A representative output from our thematic analysis process, showing how quotations from academic and grey literature were distilled into keywords, abstracted into codes, and grouped under broader themes. ( \includegraphics[height=1em]{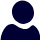}: grey literature; \includegraphics[height=1em]{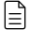}: academic literature )}
    \label{fig:code_sample}
  \Description{Illustration of qualitative coding and thematic abstraction from quotes to themes.}
\end{figure}

We applied thematic analysis to synthesize insights across our research questions, ensuring that coding remained objective and independent of the data extraction structure or biases~\cite{Stewart24}. The full papers were subsequently imported into ATLAS.ti~\footnote{https://atlasti.com/}, a qualitative data analysis tool that allowed for direct text coding. This process adhered to established thematic analysis steps~\cite{braun2006using}, providing consistency and depth in our approach.
We used a step-by-step approach to systematically identify codes, clusters, and classifications related to our research questions~\cite{kumara2021s}:

\begin{itemize}
    \item \textbf{Familiarization with the Data}: We began by reading the selected papers to understand the data comprehensively, enabling us to identify the key elements relevant to our research questions.
    \item \textbf{Pilot Study}: We randomly selected twenty sources for initial analysis, with the first author applying structural and descriptive coding~\cite{saldana2021coding} to conceptualize data relevant to the research questions~\cite{kumara2021s}. The process began by extracting quotations—direct statements from the primary studies—and identifying keywords that captured their core meaning. For example (see Figure \ref{fig:code_sample}), from quotations referring to the need to introduce failure scenarios and monitor experiment duration, the keywords \textit{Introduce failure scenarios} and \textit{Experiment duration} were extracted. These were then grouped to form the code \textit{Experiment Execution}. Similarly, the keywords \textit{Monitoring sidecar} and \textit{Observability for diagnosis} were extracted and abstracted into the code \textit{Monitoring and Observability}. These codes were clustered into the broader theme \textit{Functional Requirements}. This phase yielded 536 initial codes and an initial set of 28 candidate themes, which served as the foundation for subsequent refinement.
    
    \item \textbf{Inter-Rater Consensus Analysis for the Pilot Study}: After completing the pilot coding, the second author reviewed the full set of codes independently, assessing the naming, scope, and categorization of each code. The discrepancies between authors were discussed and resolved collaboratively. This consensus process refined the initial 536 codes to a validated set of 421 codes, improving the clarity and consistency of the thematic structure.
   
    \item \textbf{Complete Dataset Coding}: The remaining sources were subsequently coded using the validated codebook. The first author conducted the full-dataset coding, and the second author reviewed the outputs. Discrepancies identified during this phase were resolved collaboratively.

    \item \textbf{Inter-Rater Consensus Analysis for the Complete Study}: After the complete coding process, a final inter-rater assessment was conducted to ensure consistent application of the codes across the dataset. The discrepancies in interpretation were discussed and resolved. During this phase, the initial 28 themes were critically reviewed for internal coherence and conceptual distinctiveness. Overlapped or weakly supported themes were consolidated or removed, resulting in a final set of 15 distinct, non-overlapping themes. At this stage, each finalized theme was explicitly mapped to the specific research question(s) it addressed (e.g., RQ1.1 – Functional Requirements, RQ3.2 – Best Practices), ensuring that the thematic framework remained purposeful and traceable throughout the analysis.
    \item \textbf{Generating the report}: The results of this thematic analysis are presented in Sections~\ref{sec:Definition}, ~\ref{sec:solved_challenges}, ~\ref{sec:Framework}, ~\ref{sec:Current_state} and ~\ref{sec:Discussion}.
    \end{itemize}

\subsection{Replication Package}

\label{sec:appendix}
For the validation and reproducibility of this work, we have provided an extensive replication package online~\footnote{\url{https://doi.org/10.5281/zenodo.15696868}}. This includes the complete compilation of sources, search queries, and thematic analysis performed using Atlas.ti (such as codes, groupings, quotations, and analyses), as well as the associated code tables.

\section{Chaos Engineering: Definitions, Functionaries, and Elements (RQ1)}
\label{sec:Definition}
This section first reviews how chaos engineering is defined across academic and industry sources. Next, it outlines its core activities and commonly associated functionalities (RQ1.1). Finally, it introduces the essential components (RQ1.2) and quality requirements (RQ1.3) that form the foundation for implementing effective chaos engineering platforms.

\subsection{Definitions}

Chaos engineering has garnered significant attention~\cite{monroeinvestigate} in both academia and industry, yet its definition remains fragmented and inconsistent. To address this, we analyzed a set of definitions identified during our thematic analysis process. These include representative definitions from each, summarized in Tables~\ref{tab:grey_defs} and~\ref{tab:academic_defs}. Across industry and academic sources, several key elements consistently emerge: \textbf{system resilience}, \textbf{failure injection}, \textbf{controlled experiments}, and \textbf{testing in production or production-like environments}. These recurring elements highlight chaos engineering's focus on deliberately introducing failures in a controlled manner to assess how systems behave under stress and recover from disruptions.

\begin{table}[htbp]
\centering
\caption{Practitioners' Definitions of Chaos Engineering}
\label{tab:grey_defs}
\small
\begin{tabular}{|p{0.75\textwidth}|p{0.20\textwidth}|}
\hline
\textbf{Definition} & \textbf{Articles Referenced} \\
\hline

\textit{"The practice of deliberately breaking systems in controlled ways to identify potential failure points before they occur naturally."} 
& \cite{Tomka24, LTIMindtree23, Abdul2024, Green2023, Krivas2020, Vinisky2024} \\
\hline

\textit{"The methodology of creating disruptive events and stressing applications to observe system responses and implement targeted improvements."} 
& \cite{Gill21, Bhaskar22, Bairyev2023, Kadikar2023} \\
\hline

\textit{"The technique of gauging system resiliency by intentionally causing failures under load to test self-healing mechanisms."} 
& \cite{Patel22, Gill2022, Bairyev2023, Kadikar2023, Kostic2024} \\
\hline

\end{tabular}
\end{table}

\begin{table}[htbp]
\centering
\caption{Academic Definitions of Chaos Engineering}
\label{tab:academic_defs}
\small
\begin{tabular}{|p{0.75\textwidth}|p{0.20\textwidth}|}
\hline
\textbf{Definition} & \textbf{Articles Referenced} \\
\hline

\textit{"A disciplined practice of experimenting on a software or distributed system to build confidence in its ability to withstand turbulent or unexpected conditions in production."} 
& \cite{fogli2023chaos, torkura2020cloudstrike, pierce2021chaos, al2024exploring, vu2022mission, jernberg2020getting, naqvi2022evaluating, Vinisky2024, basiri2016chaos, sousa2018engineering, alvaro2016automating} \\
\hline

\textit{"An empirical and scientific approach for learning how software systems behave under stress by conducting controlled experiments to evaluate and improve their resilience and fault-handling mechanisms."} 
& \cite{zhang2019chaos, sondhi2021chaos, ikeuchi2020framework, malik2023chess} \\
\hline

\textit{"A resilience-testing approach involving the intentional injection of failures or disruptions into a system—especially in production or production-like environments—to uncover weaknesses and improve fault tolerance."} 
& \cite{poltronieri2022chaos, ji2023perfce, kesim2020identifying, konstantinou2021chaos} \\
\hline

\end{tabular}
\end{table}

We followed the synthesis method proposed by Gong and Ribiere~\cite{gong2021developing}, drawing on conceptual evaluation criteria from Wacker~\cite{wacker2004theory} and Suddaby~\cite{suddaby2010editor} (see our online appendix Section~\ref{sec:appendix} for details). A total of 48 definitions were analyzed, 27 from academic sources and 29 from grey literature. Based on this evaluation, 34 definitions were retained and 22 were excluded due to issues such as vagueness, conceptual overlap, or lack of definitional clarity. These inconsistencies highlight the difficulty of establishing a unified understanding of chaos engineering—a challenge observed in previous research, 
where fragmented definitions can hinder conceptual clarity and slow down the development of standardized practices \cite{gong2021developing,degefa2021comprehensive,diaz2024large,awasthi2025digital}. As demonstrated in those contexts, establishing a clear and shared definition can help eliminate conceptual ambiguity, enable the development of sound theory and valid constructs, and strengthen the connection between research and practice. Based on these insights, we propose the following unified definition:

\begin{quote}
        \textit{Chaos engineering is a resilience testing practice that intentionally injects controlled faults into software systems in production-like or actual production environments to simulate adverse real-world conditions. It enables uncovering hidden vulnerabilities in systems that hinder their ability to withstand potential real-world disruptions. It also enables the assessment and enhancement of systems' capacity and operational readiness against such disruptions.} 
\end{quote}

In line with the definition, chaos engineering relies on a set of core activities to guide the design and execution of experiments under controlled conditions~\cite{basiri2016chaos,Aisosa2023}. Table~\ref{tab:principle_attributes} summarizes these activities.

\begin{table}[!htbp]
\caption{Core Activities in the Chaos Engineering Lifecycle}
\label{tab:principle_attributes}
\small
\begin{tabular}{|p{0.12\textwidth}|p{0.39\textwidth}|p{0.38\textwidth}|}
\hline
\textbf{Activity} & \textbf{Description} & \textbf{Articles Referenced} \\ \hline

\textbf{Establish Steady State} & 
Understand the system's normal behavior and hypothesize how it will handle chaos (real-world disruptive incidents/failures). & 
\cite{fogli2023chaos,torkura2020cloudstrike,zhang2019chaos,dedousis2023enhancing,simonsson2021observability,Gremlin23,Bremmers21,Gill21,Wickramasinghe23,cotroneo2022thorfi,nikolaidis2021frisbee,frank2021interactive,pierce2021chaos,frank2023verifying,konstantinou2021chaos,naqvi2022evaluating,malik2023chess,bedoya2023securing,kesim2020identifying,basiri2019automating,zhang2023chaos,siwach2022evaluating,jernberg2020getting,Jakkaraju20,Bocetta19,Palani23,Gill2022,Fawcett2020,Manna2021,Bairyev2023,Sbiai2023,Green2023,Krivas2020,Kostic2024,Abdul2024,Katirtzis22,Le22,Colyer19,Patel22} \\ \hline

\textbf{Setup Monitoring Infrastructure} & 
Set up tools to monitor the system for deviations from its normal behavior during controlled experiments. & 
\cite{zhang2019chaos,simonsson2021observability,Gremlin23,Bremmers21,Patel22,Gill21,Wickramasinghe23,poltronieri2022chaos,zhang20213milebeach,nikolaidis2021frisbee,pierce2021chaos,nikolaidis2023event,malik2023chess,kesim2020identifying,siwach2022evaluating,Sbiai2023,Abdul2024,Bairyev2023,Green2023} \\ \hline

\textbf{Conduct Chaos Experiments} & 
Design and execute experiments to simulate real-world failures. & 
\cite{fogli2023chaos,torkura2020cloudstrike,zhang2019chaos,dedousis2023enhancing,simonsson2021observability,Gremlin23,Bremmers21,Gill21,pierce2021chaos,naqvi2022evaluating,malik2023chess,bedoya2023securing,kesim2020identifying,jernberg2020getting,Jakkaraju20,Palani23,Gill2022,Fawcett2020,Manna2021,Bairyev2023,Patel22,Abdul2024,Roa2022} \\ \hline

\textbf{Test and Refine} & 
Review results, apply improvements, and repeat to boost resilience. & 
\cite{zhang2019chaos,simonsson2021observability,Gremlin23,Bremmers21,Gill21,Wickramasinghe23,jernberg2020getting,Bairyev2023,Sbiai2023,Green2023,Kostic2024,Colyer19,Roa2022,basiri2019automating,torkura2020cloudstrike,fogli2023chaos,Patel22,Wu21,Tomka24} \\ \hline

\textbf{Grow Blast Radius} & 
Gradually expand experiments to test increasingly complex systems and environments. & 
\cite{torkura2020cloudstrike,zhang2019chaos,simonsson2021observability,Gremlin23,Gill21,Wickramasinghe23,pierce2021chaos,frank2023verifying,poltronieri2022chaos,ikeuchi2020framework,nikolaidis2021frisbee,naqvi2022evaluating,malik2023chess,kesim2020identifying,zhang2023chaos,siwach2022evaluating,jernberg2020getting,Jakkaraju20,Fawcett2020,Manna2021,Bairyev2023,Sbiai2023,Tomka24,Green2023,Kostic2024} \\ \hline

\textbf{Expand to Production} & 
Run experiments in production to build confidence in the system’s ability to endure adverse real-world events. & 
\cite{Nombela23,Krishnamurthy21,Moon22,simonsson2021observability,pierce2021chaos,konstantinou2021chaos,park2019simulation,Fawcett2020,willard2024fail,tucker2018business,basiri2016chaos,alvaro2016automating} \\ \hline
\end{tabular}
\end{table}

\subsection{Functionalities of Chaos Engineering (RQ1.1)}
\label{subsec:functionalties}

\begin{figure}[!htbp]
  \centering
    \vspace{-10pt}
  \includegraphics[width=0.8\textwidth]{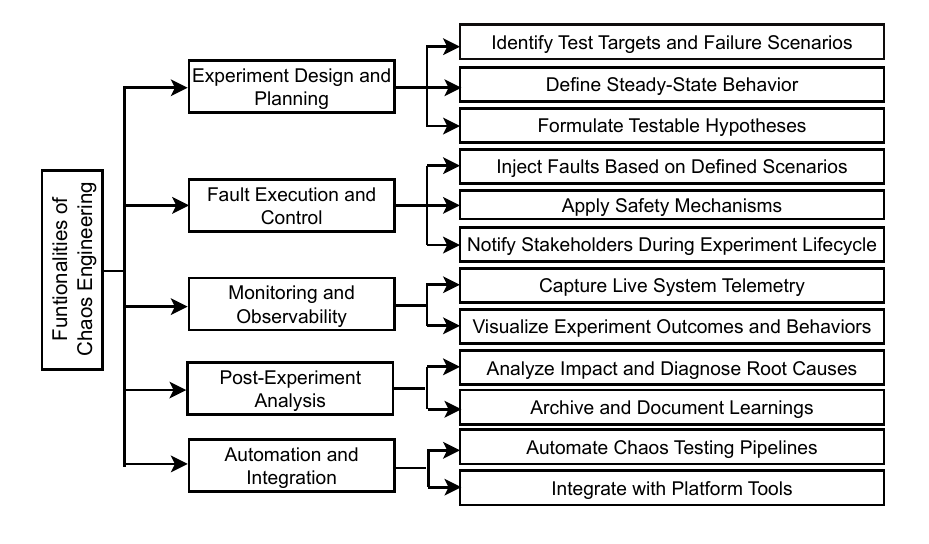}
    \vspace{-10pt}
  \caption{Key Functionalities Provided by Chaos Engineering.}
  \label{fig:CE_FUNCTIONAL_REQ}
  \Description{Functional Requirement}
\end{figure}

Chaos engineering comprises a structured set of core functionalities designed to reveal system weaknesses under real-world failure conditions. Based on our thematic analysis of academic and grey literature (see Section \ref{sec:thematic}), we identified five key functionalities that characterize how chaos engineering is typically conducted: \textit{experiment design and planning}, \textit{fault execution and control}, \textit{monitoring and observability}, \textit{post-experiment analysis}, and \textit{automation and continuous integration} (see Figure~\ref{fig:CE_FUNCTIONAL_REQ}). For full transparency, the detailed mapping of raw codes to these functionalities and supporting themes is provided in the online appendix (see Section~\ref{sec:appendix}). The remainder of this section describes each functionality, emphasizing its purpose and role in validating system resilience.

\subsubsection{Experiment Design and Planning}
This phase establishes the conceptual and procedural foundation for chaos experimentation. It ensures that testing is deliberate, measurable, and risk-aware~\cite{fogli2023chaos, simonsson2021observability,Patel22,vu2022mission}. It involves three essential activities: identifying test targets and failure scenarios, defining steady-state behavior, and formulating testable hypotheses.

\paragraph{Identify Test Targets and Failure Scenarios}
The first step is to identify components whose failure could significantly disrupt operations, such as microservices, APIs, databases, or message queues~\cite{Patel22, naqvi2022evaluating, vu2022mission, Bremmers21}. 
Failure scenarios are then defined based on system architecture analysis (e.g., service dependencies, communication patterns), operational risks, or past incident reports~\cite{fogli2023chaos, chen2024microfi, malik2023chess,nikolaidis2023event}. Typical disruptions include network latency, packet loss, service crashes, or CPU throttling~\cite{simonsson2021observability, chen2024microfi}. These scenarios are prioritized according to their likelihood and potential business impact~\cite{fogli2023chaos, naqvi2022evaluating,siwach2022evaluating,torkura2020cloudstrike}.

\paragraph{Define Steady-State Behavior}
Steady-state refers to the normal, healthy performance of the system. Teams define this using measurable indicators such as latency, error rates, throughput, or availability~\cite{frank2023verifying, malik2023chess,simonsson2021observability}. These metrics act as baselines against which system behavior is compared during and after failure injection~\cite{naqvi2022evaluating, siwach2022evaluating,Gill21}.  Establishing these baselines enables reliable detection of performance degradation and validation of resilience mechanisms~\cite{naqvi2022evaluating,Gremlin23}.

\paragraph{Formulate Testable Hypotheses}
Hypotheses describe the expected system response to injected faults. For instance, a hypothesis might state: “If service A fails, traffic will reroute to service B within 200ms without user impact”~\cite{dedousis2023enhancing, simonsson2021observability}. These statements guide the design of experiments and help interpret the results.
To manage uncertainty and uncover hidden risks, teams can use structured frameworks such as the Rumsfeld Matrix, which categorizes system knowledge as follows~\cite{Gremlin23,Bocetta19,dedousis2023enhancing,pierce2021chaos,malik2023chess,LTIMindtree23}.
\begin{itemize}
    \item \textbf{Known Knowns}: Aspects of the system that are fully understood and predictable.
    \item \textbf{Known Unknowns}: Recognized uncertainties that pose potential risks.
    \item \textbf{Unknown Knowns}: Understood risks that may not be immediately apparent or considered.
    \item \textbf{Unknown Unknowns}: Completely unforeseen issues that may emerge during experimentation.
\end{itemize}.
This framework supports a gradual testing strategy—starting with predictable behaviors and extending to more uncertain or unforeseen failure modes. Doing so helps reduce risk while expanding organizational understanding of system weaknesses~\cite{Palani23,Gremlin23,chen2022big}.

\subsubsection{Fault Execution and Control}
This phase ensures that faults are introduced in a controlled, safe, and purposeful manner to validate the system’s resilience under adverse conditions. The goal is to simulate real-world failures while minimizing risk and maximizing learning~\cite{zhang2021maximizing, Green2023,Patel22}.

\paragraph{Inject Faults Based on Defined Scenarios}
Chaos experiments begin by executing predefined fault scenarios aligned with the experiment plan. These include disruptions such as service crashes, network partitions, and CPU saturation~\cite{zhang2021maximizing, cotroneo2022thorfi,nikolaidis2023event,siwach2022evaluating}. Faults are introduced progressively—starting with a minimal scope and gradually increasing impact—to monitor system behavior and avoid widespread disruption~\cite{fogli2023chaos, Wickramasinghe23,simonsson2021observability,chen2024microfi,Green2023}. Execution readiness checks ensure that failure types are verified, system health is stable, and rollback plans are in place~\cite{zhang2019chaos, vu2022mission,chen2022big,Kadikar2023,Bairyev2023}.
\paragraph{Apply Safety Mechanisms}
To protect against unintended system degradation, safety measures are implemented. These include automatic abort thresholds that halt experiments if system performance degrades beyond acceptable levels~\cite{frank2023verifying, malik2023chess,chen2024microfi,camacho2022chaos}. Rollback protocols ensure the system can return to a known steady state, while blast radius limits prevent faults from affecting critical components~\cite{simonsson2021observability, Green2023, Aharon2024,Gremlin23,Kadikar2023}.
\paragraph{Notify Stakeholders During Experiment Lifecycle}
Transparent communication is maintained throughout the experiment. Key stakeholders are informed at defined checkpoints (start, progress, and end of experiment) to facilitate oversight, enable timely intervention, and foster cross-functional trust in the experimentation process~\cite{torkura2020cloudstrike, siwach2022evaluating, Patel22,Bairyev2023,Le22,Green2023,zhang2019chaos}.

\subsubsection{Monitoring and Observability}
This phase enables teams to understand how the system behaves under stress by continuously collecting and interpreting telemetry data during chaos experiments. Its purpose is to detect deviations, validate resilience assumptions, and guide response strategies~\cite{zhang2019chaos, camacho2022chaos, Gremlin23,nikolaidis2021frisbee}.
\paragraph{Capture Live System Telemetry}
The process begins with configuring telemetry tools to collect real-time data, such as metrics, logs, and traces~\cite{simonsson2021observability, pulcinelli2023conceptual,nikolaidis2021frisbee}. These data streams encompass both application-level indicators (e.g., service availability, request latency) and infrastructure metrics (e.g., CPU usage, memory consumption)~\cite{Gill21, chen2024microfi}. To ensure full system visibility, team members are assigned specific monitoring scopes—some focusing on business-critical services, others on backend components or network behavior~\cite{vu2022mission, nikolaidis2023event}. When unexpected behaviors occur, teams follow predefined diagnostic steps, such as analyzing error logs, correlating metrics, and reviewing recent deployment changes, to progressively isolate and identify the faulty components~\cite{ma2023phoenix, sondhi2021chaos,Patel22,chen2022big}.
\paragraph{Visualize Experiment Outcomes and Behaviors}
Collected data is visualized through dashboards that track key metrics over time and flag deviations from expected behavior~\cite{Gremlin23, Gill21,nikolaidis2021frisbee,chen2022big,Bairyev2023,poltronieri2022chaos}. These visual tools guide real-time decisions and help verify whether system behavior aligns with predefined hypotheses~\cite{zhang2019chaos, Butow2018,Bremmers21,nikolaidis2023event}. Observations are documented with relevant metadata, such as timestamps, environment context, and test conditions—to support future analysis and continuous learning~\cite{naqvi2022evaluating, sondhi2021chaos,dedousis2023enhancing,Green2023}

\subsubsection{Post-Experiment Analysis}
This phase focuses on deriving actionable insights from chaos experiments by assessing impact, diagnosing root causes, and preserving knowledge for future use~\cite{chen2022big,Bremmers21,frank2023verifying,Patel22,al2024exploring}. The process involves three key areas of functionality.
\paragraph{Analyze Impact and Diagnose Root Causes}
After fault injection, teams analyze telemetry, such as metrics, logs, and traces, to compare system performance against baseline expectations~\cite{simonsson2021observability, vu2022mission,Gremlin23,dedousis2023enhancing}. Deviations are measured to understand the extent of impact on key indicators such as latency or availability~\cite{frank2023verifying,poltronieri2022chaos,jernberg2020getting}. This helps prioritize which failures matter most. Root cause analysis then investigates how the failure unfolded, using traces and service maps to locate the origin and propagation path of faults~\cite{chen2024microfi, ma2023phoenix,zhang20213milebeach,zhang2019chaos,Bremmers21,Bairyev2023}. 
\paragraph{Archive and Document Learnings}
All findings are documented in structured postmortem reports, which capture the fault scenario, affected services, violated hypotheses, and recovery behavior~\cite{vu2022mission, Kadikar2023, Bairyev2023, jernberg2020getting, chen2022big}. These insights are stored in internal knowledge bases to support incident response and guide future experiments~\cite{zhang2019chaos, siwach2022evaluating, Green2023}. By archiving this information, teams enable iterative learning and systemic improvements in both technical and organizational resilience~\cite{Green2023,dedousis2023enhancing}.

\subsubsection{Automation and Integration}
This phase automates chaos experiments to minimize manual effort and then integrates them into delivery pipelines and platform workflows, ensuring that resilience testing becomes routine.
\paragraph{Automate Chaos Testing Pipelines}
Automation reduces manual effort by enabling chaos experiments to run consistently across different stages of the software delivery process, such as development, staging, and production~\cite{chen2022big,pierce2021chaos,bedoya2023securing}. Teams can define test parameters and execution policies in advance, ensuring that faults are injected at predictable times or in response to deployment events~\cite{Manna2021,Bairyev2023}. This supports repeatability, reduces errors, and allows for continuous resilience validation throughout the development lifecycle~\cite{Gill21,ikeuchi2020framework}.
\paragraph{Integrate with platform tools}
To embed chaos engineering into day-to-day engineering workflows, tools are integrated with existing platforms such as CI/CD (Continuous Integration and Continuous Delivery/Deployment) pipelines (e.g., Jenkins, GitHub Actions), cloud environments  (e.g., Kubernetes, AWS), and observability frameworks (e.g., Prometheus, Grafana)~\cite{simonsson2021observability, Manna2021,camacho2022chaos,park2019simulation}. This integration enables the automated execution of chaos experiments during the build, test, and deployment phases, while also ensuring that experiment telemetry is collected and visualized in real-time. As a result, teams can detect weaknesses early in the lifecycle and improve response times during failure scenarios~\cite{Gill21, siwach2022evaluating,Bairyev2023,nikolaidis2021frisbee}.

\subsection{Key Components of Chaos Engineering (RQ1.2)}
\label{sec:components}

\begin{figure}[!htbp]
  \centering
  \vspace{-10pt}
  \includegraphics[width=0.55\textwidth]{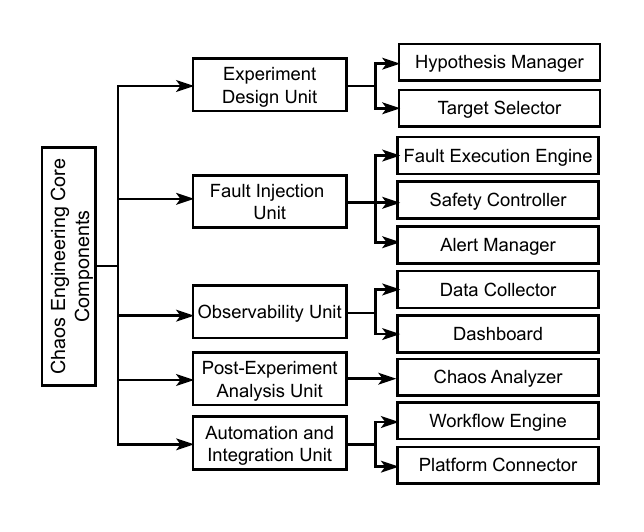}
  \vspace{-10pt}
  \caption{Key Components of Chaos Engineering.}
    \label{fig:core_components}
  \Description{Key Components of Chaos Engineering.}
\end{figure}

Before adopting a chaos engineering platform, organizations should carefully evaluate its key components to ensure that they align with the operational requirements and objectives of the organization.
Based on our systematic thematic analysis, we identified five primary components: the \textit{Experiment Design Unit}, \textit{Fault Injection Unit}, \textit{Observability Unit}, \textit{Post-Experiment Analysis Unit}, and \textit{Automation and Integration Unit} (see Figure~\ref{fig:core_components}). 
This classification is based on the functionalities discussed in Section~\ref{subsec:functionalties}, mapping each component to the corresponding phase of the experimentation lifecycle. Each unit encapsulates modular services, such as hypothesis management, failure orchestration, observability pipelines, and automation engines, that together support scalable, safe, and repeatable experimentation. The remainder of this section details each component, its sub-modules, and their roles within the broader platform architecture. For details on coding, theme development, and mapping, see our online appendix (see Section \ref{sec:appendix}).

\subsubsection{Experiment Design Unit}
This architectural component transforms high-level test objectives into system-executable configurations~\cite{soldani2021automated, zhang2021maximizing}. It offers formal interfaces to define hypotheses, select test targets, and validate readiness prior to chaos injection~\cite{Sbiai2023, chen2024microfi,dedousis2023enhancing,frank2021interactive,AzureReadiness23}.
The experiment design unit comprises two integrated modules: the Hypothesis Manager and the Target Selector.
\paragraph{Hypothesis Manager.} 
This enables teams to define, store, and manage test hypotheses through structured templates that capture expected system behavior using time-series metrics (e.g., latency, availability) and other indicators such as error rates and log patterns~\cite{Gianchandani2022,Bhaskar22,Kadikar2023,Green2023,Le22}.
Each hypothesis is programmatically linked to safety controls, including abort thresholds and rollback triggers, via policy configurations~\cite{chen2024microfi,simonsson2021observability,poltronieri2022chaos}. This ensures automated enforcement of safety during fault injection and accurate post-experiment validation~\cite{zhang2021maximizing,dedousis2023enhancing}.
\paragraph{Target Selector.}
This phase scopes where faults will be injected by allowing teams to define experiment boundaries using filters such as environment (e.g., staging, production), infrastructure layer (e.g., compute, network), and service tags or labels~\cite{soldani2021automated, AzureReadiness23,Green2023}. It ensures controlled impact by enforcing blast radius limits, so only approved targets are affected~\cite{Sbiai2023}.To maintain operational safety, it performs validation checks on all configurations, verifying correctness, authorization, and compatibility before execution. This prevents unsafe or unintended experiments from proceeding~\cite{zhang2021maximizing,cotroneo2022thorfi,Bairyev2023,chen2024microfi}.

\subsubsection{Fault Injection Unit}
\begin{figure}[!htbp]
  \centering
  \includegraphics[width=0.60\textwidth]{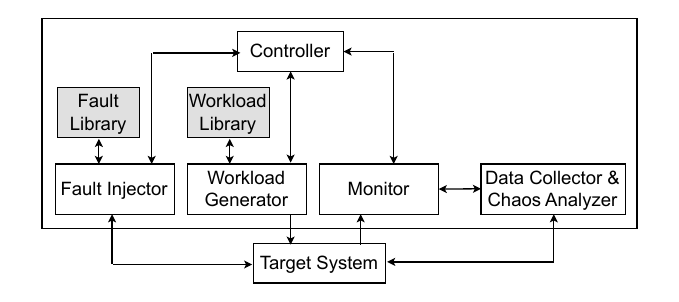}
  \vspace{-10pt}
  \caption{Conceptual Schema of Fault Execution Engine\cite{hsueh1997fault}.}
    \label{fig:fault_injection}
  \Description{Conceptual schema of fault injection.}
\end{figure}

This is the core architectural component responsible for executing chaos experiments by introducing faults into the system under test. It is composed of three core modules: \textit{the Fault Execution Engine}, \textit{the Safety Controller}, and \textit{the Alert Manager}, each supporting a distinct aspect of the fault injection lifecycle.
\paragraph{Fault Execution Engine}
This component orchestrates the actual injection of predefined faults and workloads. Its design follows the foundational model by Hsueh et al.~\cite{hsueh1997fault} (See Figure \ref{fig:fault_injection}), and has been extended to suit modern, distributed environments~\cite{soldani2021automated, frank2021interactive}. It comprises several modules:
(1) A \textbf{Controller} that sequences experiment steps and responds to telemetry, (2) A \textbf{Fault Injector} that applies disruptions from a configurable fault library,
(3) A \textbf{Workload Generator} that simulates system activity using patterns from a workload library,
(4) The \textbf{Target System}, which receives the injected faults and workloads and emits telemetry for analysis~\cite{Wickramasinghe23}. Other components from the original model (see Figure~\ref{fig:fault_injection})—such as the \textbf{Monitor}—are represented in the subsequent architectural component through the \textit{Observability Unit}, which is responsible for telemetry collection. Similarly, the \textbf{Data Collector} and \textbf{Chaos Analyzer} are addressed within the \textit{Post-Experiment Analysis Unit}, which supports impact assessment and root cause diagnosis following fault injection.
\paragraph{Safety Controller}
This enforces runtime protection by embedding safeguards directly into the fault injection lifecycle~\cite{zhang2021maximizing, cotroneo2022thorfi, Bhaskar22, lenka2018fault}. It comprises three subsystems:
\begin{itemize} 
\item \textbf{Blast Radius Control.} This limits the impact of injected faults by using traffic routing and isolation rules. For instance, in Kubernetes, these rules restrict disruptions to specific services, namespaces, or pods within the system~\cite{soldani2021automated, Wu21,vu2022mission,kesim2020identifying}.
\item \textbf{Abort Monitor.} This continuously checks key metrics (e.g., error rate, response time) and halts the experiment if thresholds are breached~\cite{Kadikar2023,Green2023,Le22}.
\item \textbf{Rollback Handler.} This restores the system to a steady state by triggering recovery workflows or replaying system snapshots~\cite{cotroneo2022thorfi, soldani2021automated,Kadikar2023,Bhaskar22}.
\end{itemize}
\paragraph{Alert Manager}
This component coordinates communication with stakeholders by triggering real-time notifications across channels (e.g., Slack, email) at critical points: start, escalation, and end~\cite{Gill21, Green2023}. It also enforces escalation policies when thresholds are breached to ensure rapid response~\cite{Patel22,AzureReadiness23,camacho2022chaos,Roa2022}.

\subsubsection{Observability Unit}
This component implements the infrastructure for capturing, processing, and presenting telemetry data during chaos experiments~\cite{pulcinelli2023conceptual, Green2023}. It is composed of two core subsystems: the \textit{Data Collector} and the \textit{Dashboard}.
\paragraph{Data Collector}
This subsystem captures diverse telemetry from systems under test~\cite{zhang2019chaos,zhang2021maximizing,soldani2021automated,naqvi2022evaluating,Mooney23}. It contains adapters/connectors for various data sources, buffering mechanisms for handling high-volume streams, and preprocessing capabilities for initial data normalization, such as timestamp alignment, unit standardization (e.g., converting memory metrics to a common format), and removal of malformed or incomplete entries~\cite{vu2022mission,nikolaidis2023event,pulcinelli2023conceptual,Green2023,camacho2022chaos}.

\textbf{Metrics Collector}
This component captures quantitative measurements, primarily in the form of time-series metrics, using instrumentation libraries and collection endpoints~\cite{nikolaidis2023event,pulcinelli2023conceptual,Green2023,AzureReadiness23}. It stores data in specialized time-series databases with retention policies and aggregation capabilities~\cite{nikolaidis2021frisbee,pulcinelli2023conceptual}. It supports both system-level metrics (CPU, memory, network) and application-level indicators (throughput, error rates) with statistical analysis functions~\cite{zhang2021maximizing,soldani2021automated,pulcinelli2023conceptual,Fawcett2020,Bhaskar22}.

\textbf{Log Aggregator}
This component collects logs from all participating systems during a chaos experiment~\cite{soldani2021automated, vu2022mission}. It uses log shippers such as Fluentd ~\footnote{\url{https://www.fluentd.org/}}  to forward entries to a central log system~\cite{Green2023,AzureReadiness23}. Logs are parsed into structured formats (e.g., JSON) for consistency~\cite{chen2022big,sousa2018engineering}. The component typically indexes log data to enable fast retrieval and supports querying by time, service, or error type~\cite{Kadikar2023,dedousis2023enhancing}.

\textbf{Tracer}
This component tracks request flows across service boundaries in distributed systems~\cite{zhang2021maximizing,nikolaidis2023event,naqvi2022evaluating}. It utilizes lightweight instrumentation, context propagation protocols, and sampling algorithms for data volume management~\cite{vu2022mission, naqvi2022evaluating, Green2023}. It can also visualize service relationships, analyze performance bottlenecks, and correlate traces with metrics and logs~\cite{naqvi2022evaluating,nikolaidis2021frisbee}.

\paragraph{Dashboard}
This component renders configurable monitoring interfaces from multiple data sources~\cite{nikolaidis2021frisbee, nikolaidis2023event, pulcinelli2023conceptual}. It can include sub-components such as query interfaces, graphical rendering engines, and templating systems~\cite{basiri2019automating, Gianchandani2022, Green2023}. The common features include real-time data refresh, threshold alerting, time-range selection, and export capabilities for sharing observations~\cite{Bhaskar22, Green2023, pulcinelli2023conceptual}.

\subsubsection{Post-Experiment Analysis Unit}
This unit implements the processes needed to evaluate chaos experiments, identify technical weaknesses, and retain organizational knowledge~\cite{zhang2019chaos, simonsson2021observability, vu2022mission, Bremmers21, Kadikar2023}. It includes two subsystems: the \textit{Chaos Analyzer} and the \textit{Knowledge Base Manager}.
\paragraph{Chaos Analyzer}
This unit processes telemetry data to assess system behavior and diagnose failure mechanisms.

\textbf{Impact Analyzer}
This module compares actual telemetry (e.g., metrics, logs, traces) against steady-state thresholds to quantify performance degradation~\cite{simonsson2021observability, vu2022mission, frank2023verifying}. It calculates impact scores based on key performance indicators (KPIs) such as latency and availability, and ranks vulnerabilities for follow-up~\cite{Bremmers21}.

\textbf{Root Cause Analyzer}
This module Correlates data across logs, traces, and metrics to trace fault propagation and uncover causal links~\cite{Kadikar2023, Kostic2024}. It applies anomaly detection and pattern-matching to distinguish root issues from surface symptoms~\cite{zhang2019chaos, simonsson2021observability}.
\paragraph{Knowledge Base Manager} This module stores post-experiment findings, including violated hypotheses, impact severity, and root causes, in structured formats. These are indexed and archived in resilience knowledge bases, supporting institutional memory, cross-team learning, and onboarding of new engineers~\cite{Manna2021, Kadikar2023, vu2022mission}.

\subsubsection{Automation and Integration Unit}
This component operationalizes the process of continuous chaos validation. It allows teams to schedule, coordinate, and trigger fault injections across environments and platforms~\cite{vu2022mission, zhang2021maximizing, Sbiai2023}. The unit includes two major subsystems: Workflow Engine and Platform Connector.
\paragraph{Workflow Engine}
This module orchestrates the automatic execution of chaos experiments based on scheduled intervals or event-driven triggers, such as deployments, pipeline stages, or config changes~\cite{vu2022mission, ma2023phoenix, basiri2019automating,fogli2023chaos,simonsson2021observability,zhang2021maximizing,cotroneo2022thorfi,camacho2022chaos,nikolaidis2021frisbee}. It applies user-defined policies to determine timing, scope, and test conditions. The engine ensures fault injection is repeatable and synchronized with real system activity, minimizing manual intervention~\cite{pierce2021chaos,nikolaidis2023event,meiklejohn2021service,chen2022big,Gill21,Bremmers21,Gill2022,Bairyev2023,Sbiai2023,torkura2020cloudstrike}
\paragraph{Platform Connector}
This facilitates seamless integration with external environments. It interfaces with CI/CD systems (e.g., Jenkins, GitHub Actions), cloud platforms (e.g., AWS, Azure, GCP), and observability tools (e.g., Prometheus, Grafana)~\cite{vu2022mission,zhang2021maximizing,frank2021interactive,nikolaidis2023event,meiklejohn2021service,malik2023chess,basiri2019automating}. Through these connections, it supports the triggering of chaos experiments based on pipeline events, deployment stages, or infrastructure changes. It also ensures synchronized data exchange across orchestration layers, telemetry systems, and alerting channels. 
\cite{fogli2023chaos,torkura2020cloudstrike,zhang2019chaos,dedousis2023enhancing,simonsson2021observability,cotroneo2022thorfi}

\subsection{Quality Requirements for Chaos Engineering Platform (RQ1.3)}
We have identified 11 key quality requirements for a chaos engineering platform from the selected literature: performance, recoverability, resilience, timeliness, observability, scalability, effectiveness, robustness, accuracy, fault tolerance, and adaptability. Table~\ref{tab:quality_metrics} provides these requirements and the corresponding metrics for measuring them.

\begin{table}[htbp]
\caption{Quality Attributes, Measurement Metrics, and Referenced Articles.}
\label{tab:quality_metrics}
\small
\begin{tabular}{|p{0.13\textwidth}|p{0.26\textwidth}|p{0.25\textwidth}|p{0.25\textwidth}|p{0.04\textwidth}|}
\hline
\makecell{\textbf{Quality} \\ \textbf{Attribute}} & \textbf{Description} & \textbf{Measurement Metrics} & \textbf{Articles Referenced} & \textbf{Total} \\\hline

\textbf{Performance} & Evaluate resource and performance under attack; align chaos tests with production metrics. & Response time (ms), CPU/memory utilization (\%), throughput (req/sec). & \cite{fogli2023chaos, torkura2020cloudstrike, zhang2019chaos, dedousis2023enhancing, Patel22, Gill21, cotroneo2022thorfi, nikolaidis2021frisbee, vu2022mission, klinaku2022beauty} & 10 \\\hline

\textbf{Recoverability} & Assess how quickly the system restores functionality after a failure. & Mean Time to Repair (MTTR), rollback time, recovery time objective (RTO). & \cite{fogli2023chaos, torkura2020cloudstrike, dedousis2023enhancing, Patel22, Gill21, cotroneo2022thorfi, camacho2022chaos, zhang20213milebeach, nikolaidis2021frisbee, pierce2021chaos, vu2022mission, frank2021interactive, malik2023chess, zhang2023chaos, Tomka24, Kadikar2023} & 16 \\\hline

\textbf{Resilience} & Evaluate the system’s ability to operate under failure without service disruption. & Percentage of successful requests under fault, error rate, degradation threshold. & \cite{fogli2023chaos, torkura2020cloudstrike, dedousis2023enhancing, Bremmers21, zhang20213milebeach, al2024exploring, vu2022mission, nikolaidis2023event, frank2023verifying, kassab2022c2b2, konstantinou2021chaos, Nombela23, Krishnamurthy21, Bocetta19, Manna2021, Kostic2024, Krivas2020, NationalAustraliaBank20, AzureReadiness23} & 18 \\\hline

\textbf{Timeliness} & Measure how quickly faults are detected and addressed. & Time to detect (TTD), alert response time. & \cite{camacho2022chaos, pierce2021chaos, vu2022mission, frank2021interactive, malik2023chess, cotroneo2022thorfi, basiri2019automating, Gill21, zhang2019chaos, Durai22, Shah21, Katirtzis22, Dua24, Wu21, Palani23, Manna2021} & 16 \\\hline

\textbf{Observability} & Assess the visibility into system internals during chaos. & Log completeness (\%), alerting coverage, trace propagation rate, telemetry accuracy. & \cite{fogli2023chaos, torkura2020cloudstrike, dedousis2023enhancing, Gremlin23, Bremmers21, Patel22, Gill21, camacho2022chaos, torkura2021continuous, zhang20213milebeach, vu2022mission, frank2023verifying, malik2023chess, kesim2020identifying, siwach2022evaluating, konstantinou2021chaos, Hawkins20, Manna2021, Kostic2024, Kadikar2023, AzureReadiness23} & 20 \\\hline

\textbf{Scalability} & Gauge how chaos testing scales with system size and complexity. & Number of nodes/services covered per experiment, test execution time at scale. & \cite{fogli2023chaos, torkura2020cloudstrike, zhang2019chaos, Patel22, Gill21, vu2022mission, sondhi2021chaos, malik2023chess, klinaku2022beauty, camilli2022microservices, nikolaidis2023event, kassab2022c2b2, rivera2023using, Krishnamurthy21, Durai22, Wu21, Palani23, Manna2021, AzureReadiness23, Vinisky2024, Kostic2024} & 20 \\\hline

\textbf{Effectiveness} & Determine whether chaos tests produce meaningful insights. & Number of unknown faults revealed, test success rate. & \cite{sondhi2021chaos, chen2022big, malik2023chess, Wu21} & 4 \\\hline

\textbf{Robustness} & Evaluate stability at peak load and error tolerance. & System uptime under stress, error rate at max load. & \cite{vu2022mission, ma2023phoenix, yu2021microrank, Wu21, bedoya2023securing, chen2022big, Vinisky2024} & 7 \\\hline

\textbf{Accuracy} & Assess how precisely the system recovers and localizes issues. & Post-recovery state correctness (\%), fault localization accuracy. & \cite{Green2023, simonsson2021observability, chen2022big} & 3 \\\hline

\textbf{Fault Tolerance} & Measure the success of self-healing and failover mechanisms. & Failover completion time, success rate of automatic recovery actions, redundancy coverage. & \cite{fogli2023chaos, torkura2020cloudstrike, zhang2019chaos, dedousis2023enhancing, Gill21, Wickramasinghe23, zhang20213milebeach, camacho2022chaos, poltronieri2022chaos, nikolaidis2021frisbee, ma2023phoenix, meiklejohn2021service, sondhi2021chaos, AzureReadiness23, Tomka24, Manna2021} & 15 \\\hline

\textbf{Adaptability} & Determine how easily the chaos system fits into varied environments. & Number of supported platforms, integration effort (e.g., setup time), configuration reuse rate. & \cite{vu2022mission, ma2023phoenix, yu2021microrank, Wu21} & 4 \\\hline

\end{tabular}
\end{table}

\section{Motivation Behind Chaos Engineering (RQ2)}
\label{sec:solved_challenges}

\begin{figure}[!htbp]
  \centering
  \includegraphics[width=1.0\textwidth]{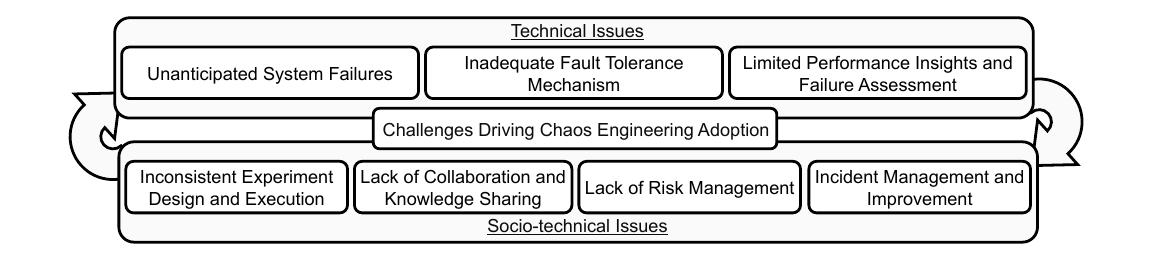}
  \caption{Challenges that Drive Chaos Engineering Adoption.}
  \label{fig:CE_Challenges_final}
  \Description{Challenges that drive chaos engineering adoption.}
\end{figure}

This section directly addresses \textbf{RQ2} by delving into the challenges driving chaos engineering adoption in organizations. Figure~\ref{fig:CE_Challenges_final} shows the technical and socio-technical challenges we identified from the reviewed literature. These challenges were identified through our coding and thematic analysis (see Section~\ref{sec:thematic}) and are documented in detail in our online appendix (see Section~\ref{sec:appendix}).

\subsection{Technical Challenges}
Chaos engineering aims to tackle various technical challenges that undermine system resilience.For example (see Section~\ref{sec:appendix}), the theme ‘Unanticipated System Failures’  includes codes such as ‘network-partition’ and ‘cascade-failure’, capturing failures that are often missed by traditional software testing. Similarly, the theme ‘Inadequate Fault Tolerance Mechanisms’ incorporates codes like ‘multi-region outage’ and ‘resource exhaustion’, highlighting gaps that chaos experiments are designed to expose. The following subsections discuss these and related technical challenges in more detail.

\paragraph{Unanticipated System Failures}
In a distributed system, one of the most critical risks is the occurrence of unexpected failures in various parts of the system~\cite{ahmed2013survey}. Given the highly interconnected nature of modern systems, a failure in one component can quickly cascade into widespread disruptions across multiple services or functions~\cite{Wickramasinghe23}. However, traditional testing methods often struggle to identify deep-rooted system vulnerabilities because they focus on known scenarios and isolated components, neglecting real-world systems' dynamic and complex interactions~\cite{soldani2021automated}.
In contrast, chaos engineering adopts a proactive approach by deliberately introducing failures, such as server crashes, network interruptions, and resource exhaustion, in environments that closely resemble production. As a result, engineers can observe how the system behaves under stress, thus revealing hidden weaknesses that traditional testing methods might overlook. By focusing on unpredictable interactions between distributed components, chaos engineering effectively uncovers vulnerabilities that arise from the complexity of the system~\cite{Bairyev2023}.
For example, chaos experiments might simulate network partitioning between microservices, which can reveal issues related to data consistency and service availability~\cite{Bairyev2023, fogli2023chaos}. These controlled disruptions allow engineers to better understand how different system components interact under turbulent conditions~\cite{torkura2020cloudstrike}.

\paragraph{Inadequate Fault Tolerance Mechanisms}

Many distributed systems are built with fault tolerance mechanisms to handle routine issues, such as server failures or network delays~\cite{Gremlin23}. However, when larger-scale failures occur, e.g., multi-region outages or prolonged network disruptions, these mechanisms often fail to maintain system stability~\cite{Palani23, Silveira23, Nallainathan23}. This shortfall occurs because mechanisms designed to handle more minor issues may not be able to handle the broader impact of significant failures, leading to severe performance degradation or even total system outages~\cite{basiri2019automating, Gill21}.
Chaos engineering addresses these shortcomings by targeting and testing fault tolerance mechanisms in scenarios replicating the most significant risks to system stability~\cite{Tivey19}. For example, chaos engineering can simulate the unavailability of critical services in one region to determine whether traffic is adequately redirected to the other areas or if the load balancers of the system can effectively distribute incoming requests between redundant resources~\cite{Bremmers21, Aharon2024}. Furthermore, by testing the ability of backup systems to maintain service levels during prolonged outages, chaos engineering reveals whether these fallback mechanisms are sufficient for extended failures~\cite{Gremlin23, Wickramasinghe23,basiri2019automating}. These focused experiments help ensure systems can handle severe disruptions without experiencing widespread failure~\cite{Bairyev2023}.
\paragraph{Limited Performance Insights and Failure Assessment}
Another significant challenge in distributed systems is understanding how they perform during high-stress or failure scenarios~\cite{chen2006overcoming,fidge1996fundamentals}. When systems operate under normal conditions, traffic bottlenecks, inefficient resource usage, or service communication failures often remain hidden~\cite{fogli2023chaos}. 
This lack of visibility leaves teams unprepared for unexpected disruptions, which can lead to outages or performance degradation when systems are stressed~\cite{konstantinou2021chaos}. Chaos engineering addresses this by introducing specific disruptions, such as intentionally increasing network latency or restricting memory resources on critical components~\cite{siwach2022evaluating}. 
These controlled failures are monitored through observability tools that provide real-time visibility into metrics such as latency, error rates, and resource consumption~\cite{Wachtel23}. By analyzing these metrics, teams gain detailed insight into how services handle stress - uncovering issues such as delayed responses or inefficient resource utilization that would otherwise remain hidden during normal operations~\cite{Bremmers21, Palani23}.
\subsection{Socio-Technical Challenges}
Chaos engineering can address various socio-technical challenges, expanding its impact beyond technical issues to encompass the complex interplay between people, processes, and technology~\cite{miles2019learning,thomas2019resilience,patriarca2018resilience}.  For instance, the theme ‘Inconsistent Experiment Design’ maps to codes such as ‘variable test timing’ and ‘lack of standardization’. Additional examples and detailed code-theme mappings can be found in the online appendix(see Section~\ref{sec:appendix}).

\paragraph{Inconsistent Experiment Design and Execution}
When experiment design and execution lack consistency, teams struggle to accurately assess the behavior of the system under stress~\cite{farias2012evaluating,caferra2023dynamic}. Factors such as varying test timing, failure scenarios, or changing system states lead to unreliable data. This makes it difficult to identify critical vulnerabilities or adequately assess the system's ability to handle disruptions~\cite{klinaku2022beauty}. Without standardization, essential insights into system responses may be overlooked, creating false confidence in its resilience~\cite{danilowicz2003consensus}. Chaos engineering addresses this by introducing a structured and repeatable approach, ensuring that predefined failure scenarios such as network issues or resource depletion are applied consistently across tests~\cite{zhang20213milebeach,naqvi2022evaluating, Kostic2024,vu2022mission}. This uniformity eliminates variability, producing reliable, comparable data that allows teams to pinpoint weaknesses and confidently evaluate recovery mechanisms~\cite{Kadikar2023, AzureReadiness23}.
\paragraph{Lack of Collaboration and Knowledge Sharing}
Limited collaboration and challenges in knowledge sharing between teams, such as development, operations, and security, can lead to communication gaps and the formation of silos, where vital information may not be effectively exchanged~\cite{habeh2021knowledge,riege2005three}. This results in misaligned responses and repeated errors during system failures, which ultimately slows recovery efforts and misses opportunities for improvement~\cite{de2016collaboration}. Chaos engineering tackles this issue by fostering cross-functional collaboration, engaging all relevant teams in both the planning and execution of chaos experiments~\cite{Tomka24, Tivey19, AzureReadiness23}. For example, when simulating a specific failure, such as a service outage or database failure, development teams define the expected behavior of the system, operations teams monitor the system’s performance under stress, and security teams evaluate potential vulnerabilities exposed during the failure~\cite{Krishnamurthy21}. This multi-team involvement ensures that the insights from the experiment, whether an unexpected failure response or a missed alert, are immediately shared and understood by everyone~\cite{Green2023, Katirtzis22, Dua24}. Furthermore, the results of each chaos experiment are documented and discussed in collaborative postmortem sessions, where all teams contribute to analyzing what went wrong, how to fix it, and how to prevent similar issues in the future~\cite{siwach2022evaluating}. 
\paragraph{Lack of Risk Management}
Effective risk management is critical in both production and non-production environments, yet many systems are exposed to unexpected failures due to insufficient preparation~\cite{Bremmers21,naqvi2022evaluating}. In non-production settings, such as development or staging, failure scenarios are often inadequately tested, leading to a false sense of security~\cite{Wickramasinghe23,jernberg2020getting}. These environments typically lack the complexity of live production systems, meaning that when software is deployed, it may not be fully equipped to handle adverse real-world events~\cite{Green2023}. Consequently, when systems reach production, where they face real user traffic, external dependencies, and unpredictable conditions, even minor issues can escalate to significant outages, resulting in costly downtime and degraded user experiences~\cite{Gill21,zhang2019chaos,ji2023perfce}. Chaos engineering solves these challenges by enabling controlled experimentation that tests system behavior under simulated and real-world failure conditions~\cite{torkura2020cloudstrike, Kostic2024}. In production environments, controlled blast radiuses ensure that failures are confined to isolated parts of the system, thereby preventing widespread disruption~\cite{Palani23, siwach2022evaluating, Gianchandani2022, Mooney23, Bocetta19}. Predefined abort conditions and rollback procedures allow teams to stop experiments and recover quickly if the stability of the system is threatened~\cite{Abdul2024, Sbiai2023,siwach2022evaluating}.
\paragraph{Incident Management and Improvement}
In complex, distributed systems, incident management faces significant challenges due to the unpredictability of failures and the limitations of reactive strategies~\cite{Gianchandani2022}. In these approaches, teams respond to incidents only after they occur, focusing on diagnosing and resolving issues after disruptions have already impacted operations~\cite{Bremmers21}. This often leads to prolonged downtime and repeated failures, as reactive methods limit the opportunities to address these problems proactively~\cite{Patel22}.
Chaos engineering offers a proactive approach by simulating targeted failures that stress critical parts of the system, revealing weaknesses in the incident response process~\cite{Anwar19, siwach2022evaluating}. For example, simulating a sudden network partition might show whether monitoring systems detect the issue promptly and if the system can reroute traffic as intended~\cite{Vinisky2024, Green2023, Gill21}. Similarly, inducing a Central Processing Unit (CPU) or memory exhaustion event could expose delays in triggering alerts or inefficiencies in the automated recovery process~\cite{camacho2022chaos, Kadikar2023, Palani23, Roa2022}. By creating these controlled failure scenarios, chaos engineering helps teams identify where their incident response protocols fall short, whether in monitoring, alert escalation, or failover processes~\cite{Wu21,zhang20213milebeach}. This enables teams to proactively adjust their strategies, improving their ability to handle actual incidents faster, more accurately, and with minimal disruption~\cite{Dua24,rivera2023using}.

\begin{table}[htbp]
\caption{Mapping of Core Chaos Engineering Functionalities with Organizational Benefits.}
\label{tab:core_benefits}
\small
\begin{tabular}{|p{0.15\textwidth}|p{0.21\textwidth}|p{0.21\textwidth}|p{0.25\textwidth}|p{0.05\textwidth}|}
\hline
\textbf{Core Functionality} & \textbf{Activity Performed} & \textbf{Benefits to Organization} & \textbf{Articles Referenced} & \textbf{Total} \\ \hline

Identify Test Targets and Failure Scenarios & 
Select critical components (e.g., APIs, services) and likely fault types based on architecture and risk. & 
Focuses efforts on the most vulnerable areas, improving preparedness and reducing downtime & 
\cite{Bremmers21,torkura2020cloudstrike,zhang2019chaos,dedousis2023enhancing,simonsson2021observability,Gremlin23,Gill21,Wickramasinghe23,zhang2021maximizing,cotroneo2022thorfi,poltronieri2022chaos,zhang20213milebeach,frank2021scenario,pierce2021chaos,soldani2021automated,al2024exploring,vu2022mission,nikolaidis2023event,Moon22,Anwar19,Dua24,Gill2022,Fawcett2020,Manna2021,Gianchandani2022,Abdul2024,AzureReadiness23,Roa2022,Krivas2020,Kostic2024,Vinisky2024,Colyer19,Green2023} & 33 \\ \hline

Define Steady-State Behavior & 
Establish baseline performance metrics (e.g., latency, throughput) & 
Provides a reference for detecting anomalies and validating system health under failure. & 
\cite{fogli2023chaos,torkura2020cloudstrike,zhang2019chaos,dedousis2023enhancing,simonsson2021observability,Gremlin23,Bremmers21,Patel22,Gill21,cotroneo2022thorfi,nikolaidis2021frisbee,pierce2021chaos,frank2023verifying,konstantinou2021chaos,naqvi2022evaluating,kesim2020identifying,park2019simulation,zhang2023chaos,siwach2022evaluating,jernberg2020getting,Jakkaraju20,Le22,Moon22,Bocetta19,LTIMindtree23,Gill2022,Fawcett2020,Bairyev2023,Abdul2024,Roa2022,Aharon2024,Krivas2020,Kostic2024} & 33 \\ \hline

Formulate Testable Hypotheses & 
Describe expected system responses to faults in measurable terms. & 
Guides experiment design and enables structured interpretation of results & 
\cite{fogli2023chaos,torkura2020cloudstrike,zhang2019chaos,dedousis2023enhancing,simonsson2021observability,Gremlin23,Bremmers21,Patel22,Gill21,cotroneo2022thorfi,frank2023verifying,konstantinou2021chaos,naqvi2022evaluating,kesim2020identifying,park2019simulation,zhang2023chaos,siwach2022evaluating,jernberg2020getting,Jakkaraju20,Bocetta19,LTIMindtree23,Gill2022,Fawcett2020,Bairyev2023,Abdul2024} & 25 \\ \hline

Inject Faults Based on Defined Scenarios & 
Execute fault (e.g., CPU overload, network delay) on selected systems. & 
Validates resilience strategies; exposes weak points under controlled stress & 
\cite{zhang2019chaos,zhang2021maximizing,cotroneo2022thorfi,zhang20213milebeach,nikolaidis2021frisbee,frank2021interactive,pierce2021chaos,soldani2021automated,vu2022mission,chen2024microfi,nikolaidis2023event,ma2023phoenix,meiklejohn2021service,sondhi2021chaos,ikeuchi2020framework,chen2022big,basiri2019automating,Gill2022,Gianchandani2022,Sbiai2023,Kadikar2023,Bhaskar22} & 22 \\ \hline

Apply Safety Mechanisms & 
Enforce abort conditions, rollback procedures, and limit blast radius. & 
Minimizes impact, safeguards critical systems, ensures quick recovery. & 
\cite{Bremmers21,Patel22,Gill21,Wickramasinghe23,frank2021interactive,pierce2021chaos,soldani2021automated,al2024exploring,vu2022mission,konstantinou2021chaos,ikeuchi2020framework,chen2022big,naqvi2022evaluating,Jakkaraju20,Starr22,Katirtzis22,Dua24,Tivey19,LTIMindtree23,Bairyev2023,Sbiai2023,Kadikar2023,Green2023,Vinisky2024} & 24 \\ \hline

Notify Stakeholders During Experiment Lifecycle & 
Send alerts and updates before, during, and after chaos experiments. & 
Promotes transparency, readiness, and coordinated response among teams. & 
\cite{torkura2020cloudstrike,zhang2019chaos,simonsson2021observability,Patel22,Gill21,Wickramasinghe23,zhang2021maximizing,torkura2021continuous,nikolaidis2021frisbee,pierce2021chaos,soldani2021automated,konstantinou2021chaos,siwach2022evaluating,Krishnamurthy21,Durai22,Tomka24,Dua24,Tivey19,Manna2021,Abdul2024,Green2023,Kostic2024,Vinisky2024,AzureReadiness23} & 24 \\ \hline

Capture Live System Telemetry & 
Monitor and record real-time metrics, logs, and traces during injection. & 
Enables detailed analysis of system and supports rapid anomaly detection. & 
\cite{zhang2019chaos,dedousis2023enhancing,simonsson2021observability,Gremlin23,Gill21,camacho2022chaos,poltronieri2022chaos,nikolaidis2021frisbee,vu2022mission,chen2024microfi,nikolaidis2023event,ma2023phoenix,sondhi2021chaos,pulcinelli2023conceptual,naqvi2022evaluating,zhang2023chaos,siwach2022evaluating,jernberg2020getting,Le22,Butow2018} & 20 \\ \hline

Visualize Experiment Outcomes and Behavior & 
Display telemetry on dashboards to highlight key deviations and trends. & 
Provides insights for decisions and facilitates shared understanding. & 
\cite{fogli2023chaos,torkura2020cloudstrike,zhang2019chaos,dedousis2023enhancing,simonsson2021observability,Gremlin23,Bremmers21,zhang2021maximizing,cotroneo2022thorfi,nikolaidis2021frisbee,Green2023} & 11 \\ \hline

Analyze Impact \& Root Causes & 
Compare metrics to baselines and trace root causes. & 
Identifies weaknesses, guides improvements. & 
\cite{fogli2023chaos,torkura2020cloudstrike,zhang2019chaos,dedousis2023enhancing,simonsson2021observability,Gremlin23,Bremmers21,zhang2021maximizing,cotroneo2022thorfi,poltronieri2022chaos,zhang20213milebeach,nikolaidis2021frisbee,pierce2021chaos,soldani2021automated,Bairyev2023,Green2023} & 16 \\ \hline

Archive and Document Learnings & 
Store findings, root causes, and lessons learned in accessible knowledge bases. & 
Preserves knowledge and supports future analysis, audits, and training. & 
\cite{fogli2023chaos,torkura2020cloudstrike,zhang2019chaos,dedousis2023enhancing,simonsson2021observability,Gremlin23,Bremmers21,Patel22,Gill21,al2024exploring,vu2022mission,frank2023verifying,chen2022big,kesim2020identifying,basiri2019automating,camilli2022microservices,yu2021microrank,siwach2022evaluating,konstantinou2021chaos,Wu21,Kadikar2023,Green2023} & 22 \\ \hline

Automate Chaos Testing Pipelines & 
Schedule and execute experiments with scripts or workflows. & 
Enhances consistency, lowers overhead, supports continuous validation. & 
\cite{torkura2020cloudstrike,zhang2019chaos,simonsson2021observability,Gill21,nikolaidis2021frisbee,pierce2021chaos,ikeuchi2020framework,chen2022big,bedoya2023securing,park2019simulation,siwach2022evaluating,Manna2021,Green2023} & 13 \\ \hline

Integrate with Platform Tools & 
Link chaos tools to CI/CD, monitoring, and cloud systems for orchestration. & 
Embeds resilience testing into delivery pipelines and ensures synchronized operations. & 
\cite{torkura2020cloudstrike,dedousis2023enhancing,Bremmers21,cotroneo2022thorfi,camacho2022chaos,poltronieri2022chaos,nikolaidis2021frisbee,frank2021interactive,soldani2021automated} & 9 \\ \hline

\end{tabular}
\end{table}

\subsection{Benefits}
Table~\ref{tab:core_benefits} outlines the organizational benefits derived from the functionalities described in Section~\ref{subsec:functionalties}, addressing the previously identified challenges. 
These functionalities were identified through our thematic analysis process.
They include defining chaos experiments, facilitating team communication, notifying stakeholders, maintaining documentation, and using real-time monitoring dashboards. The corresponding activities, such as establishing experimental parameters, analyzing system performance metrics, and fostering cross-functional collaboration, contribute to several organizational advantages. These include improved system understanding, enhanced error-handling capabilities, vital team coordination during stress conditions, and proactive system tuning. Furthermore, automating tasks and monitoring systems in real-time can minimize disruptions and optimize system performance and user experience.

\section{Chaos Engineering Framework: Taxonomy, Tools, Practices, and Evaluation (RQ3)}
\label{sec:Framework}
This section addresses \textbf{(RQ3)} by outlining proposed solutions in chaos engineering. It presents a taxonomy of tools and techniques \textbf{(RQ3.1)}, highlights best and bad adoption practices \textbf{(RQ3.2)}, and describes the methods used to evaluate chaos engineering approaches \textbf{(RQ3.3)}.

\subsection{Taxonomy of Chaos Engineering \textbf{(RQ3.1)}}
\begin{figure}[!htbp]
  \centering
    \vspace{-10pt}
  \includegraphics[width=0.99\textwidth]{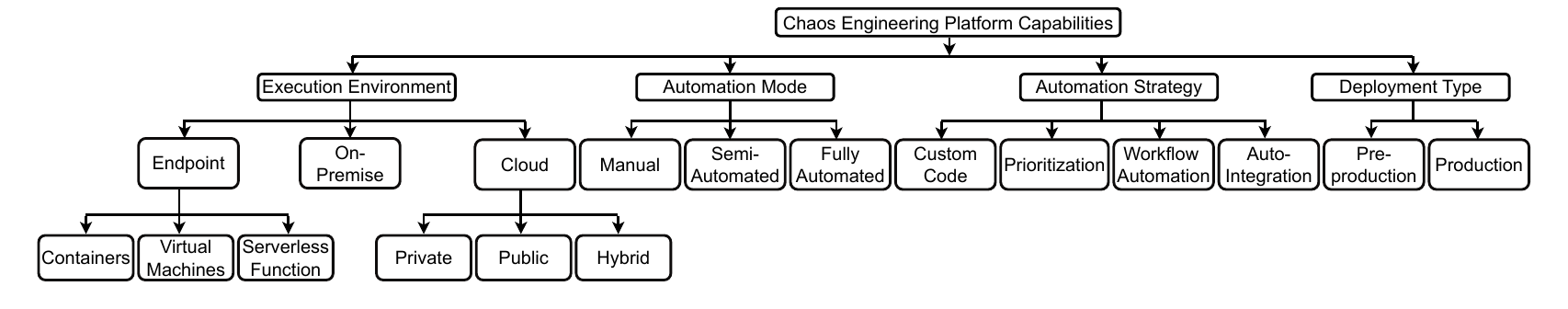}
    \vspace{-10pt}
  \caption{A Taxonomy of Chaos Engineering Platforms and Their Capabilities.}
    \label{fig:CE_TAXX}
  \Description{A taxonomy of chaos engineering platforms.}
\end{figure}

To address RQ3.1, we propose a taxonomy (Figure~\ref{fig:CE_TAXX}) that organizes chaos engineering platforms across several key dimensions: execution environment, automation strategy, automation mode, deployment type, and evaluation approach. The dimensions of this taxonomy were developed from the core themes identified through our coding and thematic analysis.
Table~\ref{tab:my-table55} presents an excerpt of the intermediate coding structure that grounds this taxonomy, mapping each conceptual category to its subcodes and supporting sources.
This taxonomy provides organizations with a structured approach to adopting chaos engineering practices by offering clear guidelines for selecting appropriate execution environments, choosing automation strategies that integrate essential tools such as monitoring, logging, and observability, and determining the optimal deployment types (e.g., production or pre-production) based on their risk tolerance. It also helps define task modes (manual, semi-automated, or fully automated).  
Additionally, it guides the selection of chaos engineering tools by aligning them with specific combinations of these dimensions—for example, tools differ in aspects such as the types of faults they inject and the level of automation they support, ensuring that organizations have the necessary evaluation methods to measure the impact of experiments and derive actionable insights for system improvement.

\begin{table}[h]
\centering
\caption{Representative Codes and Subcodes for the Taxonomy of Chaos Engineering Platforms and Their Capabilities.}
\label{tab:my-table55}
\small
\begin{tabular}{|p{2.7cm}|p{2.5cm}|p{9.2cm}|}
\hline
\textbf{Code} & \textbf{Sub-code} & \textbf{Sub-code} \\
\hline

\multirow{2}{*}{Execution Environment} & \multirow{2}{*}{Virtual Machines} & Variability introduced by different guest OS configurations. \\ \cline{3-3}
 &  & Resource contention issues examined under CPU/memory saturation scenarios. \\
\hline

\multirow{3}{*}{Automation Mode} & \multirow{3}{*}{Manual} & Tends to be time and resource intensive, as each step must be performed and monitored manually. \\ \cline{3-3}
 &  & Experiments are custom designed for specific systems or hypotheses. \\ \cline{3-3}
 &  & Carries risk of human error, especially in timing, targeting, or recovery. \\
\hline

\multirow{2}{*}{Automation Mode} & \multirow{2}{*}{Semi-Automated} & Engineer-Defined Parameters. \\ \cline{3-3}
 &  & Human-in-the-Loop Oversight. \\
\hline

\multirow{3}{*}{Automation Strategy} & \multirow{3}{*}{Custom Code} & Used to simulate specific failure types (e.g., service crash, DB outage). \\ \cline{3-3}
 &  & Custom scripts extend existing chaos tools to support advanced scenarios. \\ \cline{3-3}
 &  & Allows precise control over failure timing, scope, and recovery validation. \\
\hline

\multirow{2}{*}{Automation Strategy} & \multirow{2}{*}{Prioritization} & Critical Component Identification. \\ \cline{3-3}
 &  & Resource Allocation Strategy. \\
\hline

\multirow{2}{*}{Automation Strategy} & \multirow{2}{*}{Workflow Automation} & Ensures repeatability and reduces manual error. \\ \cline{3-3}
 &  & Supports multi-step or multi-system coordination. \\
\hline

\multirow{2}{*}{Deployment Type} & \multirow{2}{*}{Production} & Live Environment Testing. \\ \cline{3-3}
 &  & Real-User Interaction Observation. \\
\hline

\multirow{2}{*}{Deployment Type} & \multirow{2}{*}{Pre-production} & Complement to Production Testing. \\ \cline{3-3}
 &  & Staging Environment Simulation. \\
\hline

\end{tabular}%

\end{table}

\subsubsection{Execution Environment}

This refers to the underlying infrastructure and system components where chaos engineering experiments are performed. These components are categorized into endpoints, grouped by resource types and cloud environments, representing various deployment models. This section will discuss both of these in detail.

\paragraph{Endpoints}

These are physical or virtual devices connected to a computer network to exchange data. These include virtual machines (VMs), containers, and serverless functions~\cite{Microsoft2024}. Each type of endpoint presents unique challenges that chaos engineering addresses by simulating failure conditions to improve the robustness of the system and ensure continuous operation.

Virtual machines (VMs) offer a layer of abstraction over physical hardware, allowing multiple independent operating systems to share resources efficiently~\cite{sharma2016containers,smith2005architecture,basiri2016chaos}. However, VMs can experience performance issues when resources are overutilized, and they are vulnerable to hypervisor failures and migration difficulties~\cite{AzureReadiness23,fogli2023chaos,sousa2018engineering}. Chaos engineering addresses these challenges by testing VMs under stress conditions such as CPU or memory exhaustion to assess if VMs can recover from failures or seamlessly migrate workloads~\cite{Kostic2024, chen2024microfi,Green2023}. 
    
Containers, which encapsulate applications and their dependencies in lightweight environments, provide greater efficiency and portability compared to VMs~\cite{koskinen2019containers,bedoya2023securing}. However, containers face resource contention, network instability, and orchestration failures, especially in platforms such as Docker and Kubernetes~\cite{simonsson2021observability}. Chaos engineering tests on containers simulate abrupt stoppages and network disruptions, helping to evaluate containers' ability to recover within container orchestration systems while maintaining application performance ~\cite{simonsson2021observability,camacho2022chaos,Bremmers21,Kostic2024,ikeuchi2020framework,bedoya2023securing}.
    
Serverless functions allow applications to be triggered by specific events and run in isolated containers without requiring infrastructure management~\cite{castro2017serverless,taibi2020patterns,al2024exploring,Aharon2024}. While serverless architectures are scalable and cost-effective, they face challenges such as resource limitations and the reliance on third-party cloud providers~\cite{rajan2020review,van2018addressing,nookala2023serverless}. Chaos engineering helps ensure that serverless functions can handle these challenges by simulating cold starts and execution failures, verifying that functions recover quickly and maintain performance under varying loads~\cite{Starr22, jernberg2020getting, bedoya2023securing,morichetta2023intent,chatterjee2023cloud}.

\paragraph{Cloud}
Cloud environments are integral to modern Information Technology (IT) infrastructures, providing flexible and scalable resources to meet dynamic business demands~\cite{avram2014advantages,garrison2017cloud}. These environments are typically classified into three main models: public, private, and hybrid clouds~\cite{balasubramanian2012security,gorelik2013cloud}. Each cloud model has unique characteristics, risks, and opportunities, where chaos engineering can address potential failures.

Public cloud platforms, such as AWS, Microsoft Azure, and Google Cloud, offer computing resources over the Internet that are shared between multiple organizations~\cite{saraswat2020cloud,torkura2020cloudstrike}. This model enables businesses to scale their infrastructure on demand, eliminating the need for significant investments in physical hardware. Due to its cost efficiency and scalability, the public cloud has become a popular choice for many enterprises~\cite{dutta2019comparative,Jakkaraju20,Colyer19}.
However, relying on shared infrastructure introduces specific vulnerabilities. For example, a regional outage or failure in an availability zone can have widespread impacts, simultaneously affecting many customers~\cite{paavolainen2016observed}. In addition, many organizations depend on third-party services, such as application programming interfaces (APIs) or managed databases, and failures in these services can trigger further disruptions in their systems~\cite{zhou2010services,Krishnamurthy21}.
To address these risks, chaos engineering can simulate large-scale failures, such as regional outages or availability zone failures, to test how well systems can failover to other regions without significant service interruptions~\cite{fogli2023chaos,NationalAustraliaBank20}. Moreover, by injecting failures into critical third-party services, chaos engineering enables organizations to evaluate the effectiveness of their recovery strategies, ensuring that operations can continue even in the face of service disruptions~\cite{yu2021microrank,Hawkins20,Vinisky2024}. 

Private cloud environments, which provide dedicated infrastructure exclusively for a single organization, can be hosted on-premises or managed by third-party providers, offering greater control over data privacy, security, and customization~\cite{jadeja2012cloud,gastermann2015secure}. This model suits organizations with strict regulatory requirements or managing sensitive information~\cite{pearson2013privacy}. Despite their benefits, private clouds often face scalability and geographic redundancy challenges, as they typically lack the built-in redundancy of public clouds, making them more susceptible to service interruptions due to hardware failures or resource constraints~\cite{kumar2012cloud,kassab2022c2b2}. Without adequate failover mechanisms, private cloud environments may struggle to maintain operational continuity during hardware or network failures~\cite{rimal2009taxonomy,Tivey19}. Chaos engineering can be used in private clouds to simulate hardware failures, such as server crashes or network partitions, to evaluate the system's ability to recover from localized disruptions~\cite{Patel22,simonsson2021observability}. By simulating resource exhaustion, such as CPU or memory overloads, organizations can test whether their infrastructure can scale to meet unexpected surges in demand, ensuring continuous service availability under pressure~\cite{Gill21,jernberg2020getting}.

Finally, hybrid cloud environments combine the strengths of public and private clouds, allowing organizations to manage sensitive workloads on private infrastructure while leveraging the scalability of public clouds for nonsensitive or high-demand tasks~\cite{goyal2014public,poltronieri2022chaos}. This approach provides flexibility but introduces complexity in maintaining smooth integration between the two environments~\cite{lu2014development}. A key challenge of hybrid clouds is to ensure seamless communication between public and private components, as issues such as network latency, connectivity problems, and data synchronization delays can arise when workloads are distributed across both environments, potentially leading to service disruptions~\cite{annapureddy2010security,siwach2022evaluating}.
Chaos engineering can address this issue by allowing organizations to simulate network interruptions or increased latency between public and private clouds~\cite{fogli2023chaos}. These tests help to evaluate whether workloads can continue to operate despite connectivity issues~\cite{torkura2020cloudstrike,dedousis2023enhancing}. Furthermore, chaos experiments that simulate delays or failures in data replication help ensure that the system maintains data consistency and operational continuity, even when part of the infrastructure is compromised~\cite{camacho2022chaos}.

\paragraph{On-Premise}

This refers to the computing infrastructure hosted and managed within an organization's physical location rather than relying on cloud providers~\cite{willard2024fail,nikolaidis2021frisbee,vaisanen2023security}. 
These systems offer greater control over hardware, data, and network resources, essential for industries with strict regulatory requirements. However, they also face hardware failures and scaling limitations, affecting the reliability of the systems~\cite{bystrom2022comparison}.
For example, an organization managing critical on-premise data might simulate a hardware failure, such as a server or disk crash, to test how quickly backup systems respond. If synchronization delays are found, the replication process can be optimized to ensure smooth transitions and compliance with data availability standards~\cite{fogli2023chaos,dedousis2023enhancing}.
Similarly, scaling limitations can be addressed by testing how an on-premise infrastructure responds to various demand surges, such as a sudden spike in user traffic during an e-commerce company's high-demand event~\cite{AzureReadiness23, Gianchandani2022}. This helps identify bottlenecks, optimize resource allocation, and ensure that the system remains responsive under varying load conditions~\cite{naqvi2022evaluating, Wickramasinghe23,camacho2022chaos,nikolaidis2023event,nikolaidis2021frisbee}.

\subsubsection{Experimentation Modes}
Chaos engineering platforms offer three modes of experimentation, namely manual, semi-automated, and fully automated, each tailored to different levels of control, complexity, and system scale. 

Manual experimentation is most appropriate when engineers require high precision and control, as they take full responsibility for designing, configuring, and executing chaos tests~\cite{Kostic2024,torkura2020cloudstrike}. This approach is suited for organizations with little or no experience in chaos engineering or when highly specific, customized experiments are necessary~\cite{Durai22,zhang2019chaos}. For example, an engineer might manually trigger a database failure or introduce network latency to observe the system response in real time~\cite{jernberg2020getting, Bremmers21}. While this mode allows for fine-tuned control, it can be time-consuming and prone to human error, making it ideal for small-scale, customized tests or critical systems where close monitoring is essential~\cite{ikeuchi2020framework,chen2024microfi}.

Semi-automated experimentation provides a middle ground between manual control and complete automation, where engineers define the experiment parameters~\cite{jernberg2020getting,pierce2021chaos,frank2021interactive}. At the same time, the platform automates tasks such as triggering tests and collecting data, making this mode ideal for experiments that benefit from automated execution while still requiring moderate human involvement to reduce manual effort~\cite{kesim2020identifying,nikolaidis2021frisbee, Gill21}. For example, engineers might set up an experiment to test resource exhaustion on specific microservices, with the platform automatically running the test at predefined intervals, collecting performance metrics, and allowing them to focus on adjusting parameters and interpreting results~\cite{pierce2021chaos,dedousis2023enhancing}.

Fully automated experimentation suits large-scale systems that require frequent and continuous testing~\cite{simonsson2021observability,basiri2019automating,Patel22}. After engineers configure the initial setup, the chaos engineering platform takes over, seamlessly integrating experiments into the system operations, typically through the CI/CD pipelines~\cite{konstantinou2021chaos,pulcinelli2023conceptual, Manna2021, Colyer19}. For example, after a code deployment or infrastructure update, the platform can automatically simulate a service outage or introduce network disruptions, then autonomously gather performance metrics, identify system weaknesses, and generate detailed reports, requiring engineers to intervene only when necessary~\cite{dedousis2023enhancing, Bairyev2023,kesim2020identifying,cotroneo2022thorfi,nikolaidis2021frisbee}.

\subsubsection{Chaos Automation Strategy}
Chaos engineering platform employs an automation strategy to ensure consistent and repeatable experiments, reduce manual effort, and enable continuous testing~\cite{Kostic2024, Wickramasinghe23, Gremlin23}.

\paragraph{Custom Code} 
This involves developing programs or scripts to extend the capabilities of existing tools and gain precise control over system configurations and experiments~\cite{myers2000past,thum2014featureide,khankhoje2022beyond,cotroneo2022thorfi}. It allows engineers to simulate unique failure conditions, ranging from simple component failures, such as restarting a single service, to complex multi-step sequences involving multiple systems, tailored to the specific challenges of their architecture that standard tools may not address~\cite{Gill21, Gremlin23,torkura2020cloudstrike,AzureReadiness23,Patel22,nikolaidis2023event}.
For example, custom code can terminate random containers in distributed environments such as Kubernetes, simulating unexpected service failures and testing the system's recovery capacity by automatically restarting services~\cite{Bremmers21, Fawcett2020, simonsson2021observability, vu2022mission}. 
In more complex scenarios, custom code can initiate a service outage followed by a database failure or a network partition to observe how these failures impact system recovery and resilience~\cite{Wickramasinghe23, Gill21,simonsson2021observability}. Additionally, custom code can be integrated into CI/CD pipelines to simulate high resource consumption, such as CPU throttling or memory exhaustion, after each deployment, allowing early detection of performance bottlenecks before they affect production~\cite{AzureReadiness23, konstantinou2021chaos, Manna2021, Kadikar2023, Palani23}. This custom code for chaos experiments can be written in various programming languages, such as Python, Bash, or Go, depending on the complexity and specific requirements of the experiment~\cite{siwach2022evaluating, malik2023chess,sondhi2021chaos,simonsson2021observability,zhang2021maximizing}.

\paragraph{Auto-Integration of Third-Party Tools} 
This process automatically integrates external systems, such as monitoring, log, observability, alerting, incident management, and security platforms, into chaos engineering experiments~\cite{torkura2020cloudstrike,simonsson2021observability}. This automated integration ensures continuous data collection without manual intervention, enabling real-time analysis during system disruptions or failures. As a result, teams can monitor performance, identify issues quickly, and respond more efficiently to failures~\cite{Palani23, Kostic2024}.
This process is vital to integrating monitoring tools that track essential system metrics such as CPU usage, memory consumption, and network performance~\cite{Wu21, Gianchandani2022,zhang2019chaos}. Monitoring tools can automatically detect performance deviations and provide immediate feedback. For example, if a service is overloaded due to an unexpected spike in user requests, monitoring systems measure the increase in CPU usage and network load, helping the team assess the impact and determine whether the system can handle the demand or if further intervention is needed~\cite{Wickramasinghe23,fogli2023chaos,cotroneo2022thorfi, Gill21,siwach2022evaluating,malik2023chess}.
Logging tools record vital events such as service failures, error messages, and recovery attempts, documenting the system's response in real-time. For example, if network communication between two microservices fails, the logging systems capture the exact moment the failure occurred and show the error messages generated by failed retries~\cite{cotroneo2022thorfi,ma2023phoenix}. This data helps teams analyze how the failure affected communication and whether recovery mechanisms worked effectively~\cite{naqvi2022evaluating,malik2023chess, Kostic2024,Palani23}.
Building on this, observability tools provide a broader view by correlating logs and metrics to show how different components of the system interact during failure scenarios. These tools enable teams to identify how failures propagate across interconnected services~\cite{simonsson2021observability,Manna2021, Green2023,Kostic2024, Krishnamurthy21}. For example, if latency is introduced between services in a chaos experiment, observability systems track the delays at each hop between services and how this latency affects downstream components such as database queries or API responses. This tracing allows the team to see if minor delays in one service cause issues elsewhere~\cite{Durai22, Katirtzis22,Kostic2024,chen2024microfi,nikolaidis2023event,frank2023verifying}.
Furthermore, alert notification systems can ensure that relevant teams are promptly notified when critical failures occur or performance thresholds are exceeded~\cite{Tivey19, Gianchandani2022, Palani23, Bremmers21, Starr22}. For example, if a database connection pool reaches its maximum capacity during a high-load chaos test, the alert system triggers a notification to the database operations team, allowing them to address the issue before it causes a system-wide failure. This early warning helps prevent more significant disturbances~\cite{Green2023,malik2023chess}.
Finally, incident management tools streamline the resolution process in conjunction with alerting by automatically generating tickets for failures detected during chaos tests. These tickets are assigned to the appropriate teams and tracked until the issue is resolved~\cite{Gremlin23, Patel22, Green2023, Le22}. For example, if a chaos experiment causes a loss of connection to external APIs, the incident management system creates a ticket assigned to the appropriate API support team, ensuring that the problem is tracked and resolved efficiently~\cite{rivera2023using,siwach2022evaluating}.

\paragraph{Workflow}

A chaos engineering workflow can automate the controlled execution of experiment steps, such as setting up the environment, executing failures, monitoring, and cleanup, ensuring that the experiments are consistent and repeatable ~\cite{camilli2022microservices}. This reduces the risk of errors and guarantees the reliability in each run. For example, automating a database outage simulation ensures that critical steps such as downgrading the database, monitoring the system response, and restoring functionality are executed reliably every time ~\cite{Wickramasinghe23}. In addition, workflows simplify complex experiments that involve multiple systems or services. In such cases, they can coordinate the simultaneous failure of various microservices, allowing teams to test the system’s ability to maintain service availability despite several disruptions ~\cite{nikolaidis2021frisbee, Manna2021}. Furthermore, a workflow engine ensures that each step is executed correctly while capturing and analyzing the results, making the experimentation process both structured and efficient~\cite{torkura2020cloudstrike, Bairyev2023}.

\paragraph{Prioritization}
This involves categorizing system components into critical and non-critical groups to focus testing efforts on the areas where failures would have the greatest impact~\cite{dedousis2023enhancing, Anwar19,chen2024microfi}. Critical components are essential for maintaining core operations, and their failure would cause significant disruption~\cite{fogli2023chaos,kesim2020identifying}. For example, API gateways and distributed databases are critical because these failures can disrupt communication between services, cause widespread outages, or lead to data loss and downtime, directly affecting users~\cite{siwach2022evaluating}. In contrast, non-critical components have minimal immediate impact on the core functionality of the system. For instance, log aggregation systems are non-critical; their failure may affect internal monitoring but will not disrupt user-facing services or essential system functions~\cite{dedousis2023enhancing}. By prioritizing critical components, testing can focus on areas where failure would result in the most severe consequences. In contrast, the resilience issues of non-critical components can be addressed less urgently, since their failure poses a lower risk to overall system stability~\cite{Wickramasinghe23,Gremlin23}.

\subsubsection{Deployment}
The type of deployment selected for chaos experiments is a critical decision in chaos engineering, as it defines the testing environment and significantly affects the performance and stability of the system~\cite{meiklejohn2021service,Abdul2024,zhang2019chaos}. Choosing the appropriate deployment type ensures that the experiments yield meaningful insights while minimizing the risks to live operations~\cite{siwach2022evaluating, Wickramasinghe23}. This section identifies two primary deployment types: production and pre-production.

\paragraph{Production.}
This refers to running tests or experiments in a live environment where real users actively interact with the System Under Test (SUT)~\cite{zhang2019chaos, Bremmers21,vu2022mission,nikolaidis2023event,cotroneo2022thorfi}. For example, deploying a new feature on an e-commerce website during sales periods allows teams to observe user interactions and system performance under actual conditions~\cite{jernberg2020getting}. A significant benefit of production deployment is the ability to uncover vulnerabilities that might remain hidden in more isolated testing environments~\cite{Wickramasinghe23,nikolaidis2021frisbee}. For example, testing a service during peak traffic can reveal weaknesses that would not otherwise emerge, making it crucial to gain insight into how the SUT handles real-world disruptions and help identify areas for further optimization~\cite{fogli2023chaos,chen2024microfi,simonsson2021observability}. Production deployment introduces risks, as failures during testing can directly affect users and disrupt business operations~\cite{Roa2022, jernberg2020getting}. Production experiments must be carefully planned to mitigate these risks, such as scheduling tests during low-traffic periods and implementing automated rollback mechanisms to restore normal operations if problems occur~\cite{sondhi2021chaos, Bremmers21}.

\paragraph{Pre-production.} 
This environment closely mirrors the production environment, allowing teams to validate system functionality and performance before going live~\cite{naqvi2022evaluating,vu2022mission,Bremmers21}. An example of a pre-production environment is a staging environment, where the SUT can be deployed with the same configurations as production, enabling comprehensive checks on how it will perform under expected loads~\cite{jernberg2020getting, Manna2021}. Pre-production deployment allows for complex tests that might be too risky to execute in live environments~\cite{Bhaskar22}. For example, simulating a network partition can reveal how well the SUT responds to such failures, helping teams resolve potential problems before they reach production~\cite{Green2023}. This added layer of testing ensures major releases and helps reduce the risk of failure once the system is live ~\cite{Kostic2024}. One limitation of pre-production testing is that it may not accurately reflect the complexities of real-world operations~\cite{Bhaskar22,naqvi2022evaluating}. Consequently, these tests should be viewed as a complement, rather than a substitute, to production testing~\cite{zhang2019chaos, torkura2020cloudstrike, Gianchandani2022}.

\subsection{Chaos Engineering Tools}
\label{sec:Tools}

To further address \textbf{RQ3.1}, we applied the taxonomy developed in this study to evaluate ten widely used chaos engineering tools (T1–T10 in Table~\ref{tab:Tax}).

We identified 71 tools mentioned across academic and grey sources during our thematic analysis. We filtered this list by inspecting their GitHub repositories. We excluded tools that had been archived (read-only), were no longer maintained, or lacked sufficient documentation. This screening process reduced the set to 41 active tools with publicly available repositories. From this refined pool of 41 tools, we selected the ten most starred repositories on GitHub as of November 2024. The ‘most stars’ metric was used as a proxy for community interest and real-world relevance ~\cite{borges2016understanding}. Each selected tool was evaluated using our proposed taxonomy, which classifies tools based on four key dimensions: execution environments, automation modes, failure injection strategies, and deployment stages. This mapping is presented in Table~\ref{tab:Tax} and is intended to help engineering teams align tool selection with the architecture of the systems under test(e.g., microservices, containerized environments) and operational goals (e.g.,reducing latency, enhancing scalability)~\cite{kavaler2019tool}.

For example, T1 and T2 support Kubernetes environments. T1 enables manual testing in both pre-production and production stages, while T2 extends automation with custom code and integration, targeting pre-production use. T3 to T5 offer a broader platform support, including Kubernetes, Docker, and VMs, and provide full automation features such as workflow integration, prioritization, and custom code, applicable across all deployment stages. T6 and T7 focus on Kubernetes and containers. Both support manual, semi-automated, and fully automated modes; T6 emphasizes hybrid setups with prioritization, while T7 supports Docker and targets production-grade testing. T8 and T9 operate across Kubernetes and virtual machines. T8 supports complete automation and is usable in both staging and live systems, while T9 focuses on pre-production with advanced integration options. Finally, T10 is tailored for hybrid cloud environments, supporting Kubernetes and providing full automation across both pre-production and production environments.

\begin{table}[htbp]
\centering
\caption{Top-10 Most Used Chaos Engineering Platforms.}
\label{tab:Tax}
\resizebox{\columnwidth}{!}{%
\begin{tabular}{|l|l|llllllll|lll|llll|ll|}
\hline
\multirow{3}{*}{Tool Code} &
  \multirow{3}{*}{Tool Name} &
  \multicolumn{8}{c|}{Execution Environment} &
  \multicolumn{3}{c|}{Automation Mode} &
  \multicolumn{4}{c|}{Automation Strategy} &
  \multicolumn{2}{c|}{Deployment Stages} \\ \cline{3-19} 
 &
   &
  \multicolumn{4}{c|}{Endpoint} &
  \multicolumn{1}{c|}{\multirow{2}{*}{\begin{tabular}[c]{@{}c@{}}On-\\ Premise\end{tabular}}} &
  \multicolumn{3}{c|}{Cloud} &
  \multicolumn{1}{l|}{\multirow{2}{*}{Manual}} &
  \multicolumn{1}{c|}{\multirow{2}{*}{\begin{tabular}[c]{@{}c@{}}Semi-\\ Automated\end{tabular}}} &
  \multicolumn{1}{c|}{\multirow{2}{*}{\begin{tabular}[c]{@{}c@{}}Fully \\ Automated\end{tabular}}} &
  \multicolumn{1}{l|}{\multirow{2}{*}{Prioritization}} &
  \multicolumn{1}{c|}{\multirow{2}{*}{\begin{tabular}[c]{@{}c@{}}Workflow\\ Automation\end{tabular}}} &
  \multicolumn{1}{c|}{\multirow{2}{*}{\begin{tabular}[c]{@{}c@{}}Custom \\ Code\end{tabular}}} &
  \multicolumn{1}{c|}{\multirow{2}{*}{\begin{tabular}[c]{@{}c@{}}Auto-\\ Integration\end{tabular}}} &
  \multicolumn{1}{c|}{\multirow{2}{*}{\begin{tabular}[c]{@{}c@{}}Pre-\\ production\end{tabular}}} &
  \multirow{2}{*}{Production} \\ \cline{3-6} \cline{8-10}
 &
   &
  \multicolumn{1}{l|}{Docker} &
  \multicolumn{1}{l|}{Kubernetes} &
  \multicolumn{1}{c|}{\begin{tabular}[c]{@{}c@{}}Virtual\\ Machine\end{tabular}} &
  \multicolumn{1}{l|}{\begin{tabular}[c]{@{}l@{}}Serverless \\ Function\end{tabular}} &
  \multicolumn{1}{c|}{} &
  \multicolumn{1}{l|}{Private} &
  \multicolumn{1}{l|}{Public} &
  Hybrid &
  \multicolumn{1}{l|}{} &
  \multicolumn{1}{c|}{} &
  \multicolumn{1}{c|}{} &
  \multicolumn{1}{l|}{} &
  \multicolumn{1}{c|}{} &
  \multicolumn{1}{c|}{} &
  \multicolumn{1}{c|}{} &
  \multicolumn{1}{c|}{} &
   \\ \hline
T1 & Chaos Monkey
   &
  \multicolumn{1}{l|}{} &
  \multicolumn{1}{l|}{\ding{51}} &
  \multicolumn{1}{l|}{\ding{51}} &
  \multicolumn{1}{l|}{} &
  \multicolumn{1}{l|}{} &
  \multicolumn{1}{l|}{\ding{51}} &
  \multicolumn{1}{l|}{\ding{51}} &
   &
  \multicolumn{1}{l|}{\ding{51}} &
  \multicolumn{1}{l|}{} &
   &
  \multicolumn{1}{l|}{\ding{51}} &
  \multicolumn{1}{l|}{\ding{51}} &
  \multicolumn{1}{l|}{\ding{51}} &
   &
  \multicolumn{1}{l|}{\ding{51}} &
  \ding{51} \\ \hline
T2 &
  Toxiproxy &
  \multicolumn{1}{l|}{\ding{51}} &
  \multicolumn{1}{l|}{\ding{51}} &
  \multicolumn{1}{l|}{} &
  \multicolumn{1}{l|}{} &
  \multicolumn{1}{l|}{} &
  \multicolumn{1}{l|}{} &
  \multicolumn{1}{l|}{}&
   &
  \multicolumn{1}{l|}{\ding{51}} &
  \multicolumn{1}{l|}{\ding{51}} &
  \ding{51} &
  \multicolumn{1}{l|}{} &
  \multicolumn{1}{l|}{\ding{51}} &
  \multicolumn{1}{l|}{\ding{51}} &
   \ding{51}&
  \multicolumn{1}{l|}{\ding{51}} &
  {} \\ \hline
T3 &
  Chaos Mesh&
  \multicolumn{1}{l|}{} &
  \multicolumn{1}{l|}{\ding{51}} &
  \multicolumn{1}{l|}{} &
  \multicolumn{1}{l|}{} &
  \multicolumn{1}{l|}{\ding{51}} &
  \multicolumn{1}{l|}{\ding{51}} &
  \multicolumn{1}{l|}{\ding{51}} &
   &
  \multicolumn{1}{l|}{\ding{51}} &
  \multicolumn{1}{l|}{} &
  \ding{51} &
  \multicolumn{1}{l|}{\ding{51}} &
  \multicolumn{1}{l|}{\ding{51}} &
  \multicolumn{1}{l|}{\ding{51}} &
   \ding{51} &
  \multicolumn{1}{l|}{\ding{51}} &
  \ding{51} \\ \hline
T4 & ChaosBlade 
   &
  \multicolumn{1}{l|}{\ding{51}} &
  \multicolumn{1}{l|}{\ding{51}} &
  \multicolumn{1}{l|}{\ding{51}} &
  \multicolumn{1}{l|}{} &
  \multicolumn{1}{l|}{\ding{51}} &
  \multicolumn{1}{l|}{} &
  \multicolumn{1}{l|}{\ding{51}} &
  \ding{51} &
  \multicolumn{1}{l|}{\ding{51}} &
  \multicolumn{1}{l|}{\ding{51}} &
  \ding{51} &
  \multicolumn{1}{l|}{\ding{51}} &
  \multicolumn{1}{l|}{} &
  \multicolumn{1}{l|}{\ding{51}} &
   &
  \multicolumn{1}{l|}{\ding{51}} &
  \ding{51} \\ \hline
T5 & LitmusChaos 
   &
  \multicolumn{1}{l|}{} &
  \multicolumn{1}{l|}{\ding{51}} &
  \multicolumn{1}{l|}{\ding{51}} &
  \multicolumn{1}{l|}{} &
  \multicolumn{1}{l|}{} &
  \multicolumn{1}{l|}{\ding{51}} &
  \multicolumn{1}{l|}{\ding{51}} &
  {} &
  \multicolumn{1}{l|}{\ding{51}} &
  \multicolumn{1}{l|}{\ding{51}} &
  \ding{51} &
  \multicolumn{1}{l|}{\ding{51}} &
  \multicolumn{1}{l|}{\ding{51}} &
  \multicolumn{1}{l|}{\ding{51}} &
  \ding{51} &
  \multicolumn{1}{l|}{\ding{51}} &
  \ding{51} \\ \hline
T6 &
  Kube-monkey &
  \multicolumn{1}{l|}{} &
  \multicolumn{1}{l|}{\ding{51}} &
  \multicolumn{1}{l|}{} &
  \multicolumn{1}{l|}{} &
  \multicolumn{1}{l|}{\ding{51}} &
  \multicolumn{1}{l|}{\ding{51}} &
  \multicolumn{1}{l|}{} &
  {} &
  \multicolumn{1}{l|}{\ding{51}} &
  \multicolumn{1}{l|}{\ding{51}} &
  \ding{51} &
  \multicolumn{1}{l|}{\ding{51}} &
  \multicolumn{1}{l|}{} &
  \multicolumn{1}{l|}{\ding{51}} &
  {} &
  \multicolumn{1}{l|}{\ding{51}} &
  \ding{51} \\ \hline
T7 & Pumba
   &
  \multicolumn{1}{l|}{\ding{51}} &
  \multicolumn{1}{l|}{\ding{51}} &
  \multicolumn{1}{l|}{} &
  \multicolumn{1}{l|}{} &
  \multicolumn{1}{l|}{\ding{51}} &
  \multicolumn{1}{l|}{} &
  \multicolumn{1}{l|}{\ding{51}} &
  \ding{51} &
  \multicolumn{1}{l|}{\ding{51}} &
  \multicolumn{1}{l|}{} &
  {} &
  \multicolumn{1}{l|}{} &
  \multicolumn{1}{l|}{} &
  \multicolumn{1}{l|}{\ding{51}} &
  \ding{51} &
  \multicolumn{1}{l|}{\ding{51}} &
  \ding{51} \\ \hline
T8 &
  Chaos Toolkit &
  \multicolumn{1}{l|}{} &
  \multicolumn{1}{l|}{\ding{51}} &
  \multicolumn{1}{l|}{\ding{51}}  &
  \multicolumn{1}{l|}{} &
  \multicolumn{1}{l|}{\ding{51}} &
  \multicolumn{1}{l|}{}  &
  \multicolumn{1}{l|}{\ding{51}} &
  \ding{51}  &
  \multicolumn{1}{l|}{\ding{51}} &
  \multicolumn{1}{l|}{\ding{51}} &
  \ding{51} &
  \multicolumn{1}{l|}{\ding{51}} &
  \multicolumn{1}{l|}{} &
  \multicolumn{1}{l|}{\ding{51}} &
  {} &
  \multicolumn{1}{l|}{\ding{51}} &
  \ding{51} \\ \hline
T9 &
  Powerfulseal &
  \multicolumn{1}{l|}{} &
  \multicolumn{1}{l|}{\ding{51}} &
  \multicolumn{1}{l|}{\ding{51}} &
  \multicolumn{1}{l|}{} &
  \multicolumn{1}{l|}{\ding{51}} &
  \multicolumn{1}{l|}{\ding{51}} &
  \multicolumn{1}{l|}{\ding{51}} &
  {} &
  \multicolumn{1}{l|}{\ding{51}} &
  \multicolumn{1}{l|}{\ding{51}} &
  \ding{51} &
  \multicolumn{1}{l|}{\ding{51}} &
  \multicolumn{1}{l|}{} &
  \multicolumn{1}{l|}{\ding{51}} &
  \ding{51} &
  \multicolumn{1}{l|}{\ding{51}} &
  {} \\ \hline
T10 &
  Chaoskube &
  \multicolumn{1}{l|}{} &
  \multicolumn{1}{l|}{\ding{51}} &
  \multicolumn{1}{l|}{} &
  \multicolumn{1}{l|}{} &
  \multicolumn{1}{l|}{} &
  \multicolumn{1}{l|}{\ding{51}} &
  \multicolumn{1}{l|}{\ding{51}} &
  \ding{51} &
  \multicolumn{1}{l|}{\ding{51}} &
  \multicolumn{1}{l|}{\ding{51}} &
  \ding{51} &
  \multicolumn{1}{l|}{\ding{51}} &
  \multicolumn{1}{l|}{\ding{51}} &
  \multicolumn{1}{l|}{} &
  {\ding{51}} &
  \multicolumn{1}{l|}{\ding{51}} &
  \ding{51} \\ \hline
\end{tabular}%
}

\end{table}

\subsection{Adoption Practices \textbf{(RQ3.2)}}
To address \textbf{RQ3.2}, we highlight best practices and common pitfalls in adopting chaos engineering derived from our literature sources. Successfully adopting chaos engineering requires a structured approach that balances best practices with avoiding common mistakes. To minimize risk during experimentation, organizations should follow well-defined guidelines for planning, executing, and evaluating chaos experiments ~\cite{siwach2022evaluating, Green2023}. Table~\ref{tab:best_practices_metrics}  summarizes essential best practices, such as starting with small, controlled experiments, setting clear objectives, automating tests, and closely monitoring system behavior. Additionally, organizations should limit the scope of experiments, encourage cross-team collaboration, and continuously refine processes based on experiment outcomes.
While following best practices is crucial, organizations must also avoid common pitfalls, such as those outlined in Table~\ref{tab:bad_practices_metrics}, which include failing to tailor experiments to specific systems, neglecting proper planning and hypothesis formation, and lacking adequate safety mechanisms. By adhering to best practices and avoiding these common pitfalls, organizations can confidently integrate chaos engineering into their systems and achieve successful outcomes~\cite{siwach2022evaluating}. For full details, see the code-to-theme mapping in the online appendix (Section~\ref{sec:appendix}).

\begin{table}[htbp]
\caption{Best Practices.}
\label{tab:best_practices_metrics}
\small
\begin{tabular}{|p{0.03\textwidth}|p{0.17\textwidth}|p{0.31\textwidth}|p{0.30\textwidth}|p{0.05\textwidth}|}
\hline
\textbf{No} & \textbf{Best Practices} & \textbf{Description} & \textbf{Articles} & \textbf{Total} \\ \hline

\textbf{1} & Start Small and Scale Up & 
Start with manageable tests, then expand gradually. & 
\cite{fogli2023chaos,torkura2020cloudstrike,Bremmers21,Patel22, Gill21, chen2022big, siwach2022evaluating, jernberg2020getting, Anwar19, Palani23, Kadikar2023,Green2023} & 12 \\ \hline

\textbf{2} & Define Clear Objectives & 
Set precise goals to guide each experiment’s purpose and outcomes. & 
\cite{torkura2020cloudstrike, dedousis2023enhancing, Gremlin23, Patel22, zhang20213milebeach, al2024exploring, kesim2020identifying, soldani2021automated, siwach2022evaluating, jernberg2020getting, Jakkaraju20, Dua24, Kadikar2023, Green2023} & 14 \\ \hline

\textbf{3} & Automate Experiments & 
Use automation tools to run chaos experiments regularly, ensuring consistency and frequency. & 
\cite{torkura2020cloudstrike, zhang2019chaos, Gremlin23, Gill21, Wickramasinghe23, camacho2022chaos, zhang20213milebeach, al2024exploring, vu2022mission, kassab2022c2b2, ikeuchi2020framework, naqvi2022evaluating, malik2023chess, soldani2021automated, siwach2022evaluating, jernberg2020getting, Bhaskar22, Gill2022, Fawcett2020, Bairyev2023, Sbiai2023, AzureReadiness23, Roa2022, Green2023} & 24 \\ \hline

\textbf{4} & Track and Measure Results & 
Monitor the system's performance during experiments and collect data to see the impact. & 
\cite{fogli2023chaos,torkura2020cloudstrike,dedousis2023enhancing, Gremlin23, Bremmers21, Patel22, Gill21, camacho2022chaos, torkura2021continuous, zhang20213milebeach, vu2022mission, kassab2022c2b2, ikeuchi2020framework, malik2023chess, soldani2021automated, siwach2022evaluating, konstantinou2021chaos, jernberg2020getting, Hawkins20, Manna2021, AzureReadiness23, Kadikar2023, Green2023} & 23 \\ \hline

\textbf{5} & Involve all Key Stakeholders & 
Include team members to ensure everyone is informed and aligned. & 
\cite{torkura2020cloudstrike, Gremlin23, Bremmers21, Patel22, Gill21, frank2021interactive, siwach2022evaluating, Dua24, Kadikar2023, Green2023, Gill2022,Kostic2024} & 12 \\ \hline

\textbf{6} & Document and Improve & 
Record experiments to continually improve the system. & 
\cite{torkura2020cloudstrike, zhang2019chaos, Gremlin23, Bremmers21, Patel22, Gill21, siwach2022evaluating, Palani23, Kadikar2023, Green2023,kesim2020identifying} & 11 \\ \hline

\end{tabular}
\end{table}

\begin{table}[htbp]
\caption{Bad Practices.}
\label{tab:bad_practices_metrics}
\small
\begin{tabular}{|p{0.03\textwidth}|p{0.15\textwidth}|p{0.30\textwidth}|p{0.30\textwidth}|p{0.05\textwidth}|}
\hline
\textbf{No} & \textbf{Bad Practices} & \textbf{Description} & \textbf{Articles} & \textbf{Total} \\ \hline

\textbf{1} & Misaligned Chaos & 
Chaos experiments that don’t match the system yield misleading results. & 
\cite{fogli2023chaos, torkura2020cloudstrike, zhang2019chaos, dedousis2023enhancing, vu2022mission, nikolaidis2023event, ma2023phoenix, sondhi2021chaos, jernberg2020getting, konstantinou2021chaos, park2019simulation, zhang2023chaos, rivera2023using, Patel22, Le22, Lardo19,Hochstein2016,basiri2016chaos} & 18 \\ \hline

\textbf{2} & Inadequate Experimentation & 
Without planning, clear goals, and baseline tracking, assessing chaos test impacts becomes difficult. & 
\cite{Gremlin23, Wickramasinghe23, Gill21, Bremmers21, Jakkaraju20, Le22, Hawkins20, Anwar19, Dua24, NationalAustraliaBank20, Mooney23, Starr22, Palani23, Gill2022, Fawcett2020, Gianchandani2022, Bairyev2023, Abdul2024, AzureReadiness23, Kadikar2023, Green2023, Aharon2024, Krivas2020, Kostic2024, frank2021interactive, jernberg2020getting, konstantinou2021chaos, ji2023perfce, soldani2021automated, park2019simulation, Wachtel23,alvaro2017abstracting,sousa2018engineering} & 33 \\ \hline

\textbf{3} & Safety Mechanism Oversight & 
Not testing rollback and safety features risks uncontrolled failures. & 
\cite{Gremlin23, Wickramasinghe23, Gill21, Bremmers21, Patel22, Jakkaraju20, Bhaskar22, Katirtzis22, Dua24, Starr22, Tivey19, Palani23, Bairyev2023, Sbiai2023, Kadikar2023, Green2023, Vinisky2024, pierce2021chaos, al2024exploring, vu2022mission, ikeuchi2020framework, chen2022big, naqvi2022evaluating, soldani2021automated, konstantinou2021chaos,alvaro2017abstracting,sousa2018engineering} & 27 \\ \hline

\textbf{4} & Insufficient Tooling & 
Lacking observability, automation, and integration tools makes monitoring and analyzing chaos experiments difficult. & 
\cite{Gremlin23, Wickramasinghe23, Gill21, Hawkins20, Starr22, NationalAustraliaBank20, Tomka24, Dua24, Mooney23, Lardo19, Gianchandani2022, Abdul2024, AzureReadiness23, Roa2022, Kadikar2023, Green2023, Krivas2020, Kostic2024, simonsson2021observability, nikolaidis2021frisbee, soldani2021automated, al2024exploring, vu2022mission, naqvi2022evaluating, malik2023chess, camilli2022microservices, Bremmers21, Patel22, Le22, zhang20213milebeach, cotroneo2022thorfi, ikeuchi2020framework, bedoya2023securing, basiri2019automating, pulcinelli2023conceptual, jernberg2020getting, Anwar19, Durai22, Jakkaraju20} & 39 \\ \hline

\textbf{5} & Poor Stakeholder Collaboration & 
Excluding key team members causes misalignment and resistance. & 
\cite{Bremmers21, Starr22, Le22, Krishnamurthy21, Katirtzis22, Durai22, Tivey19, Gill2022, Green2023, poltronieri2022chaos, frank2021interactive, al2024exploring, jernberg2020getting, kesim2020identifying} & 14 \\ \hline

\end{tabular}
\end{table}

\subsection{Evaluation Approaches \textbf{(RQ3.3)}}
Evaluation refers to determining the effectiveness of chaos engineering experiments by measuring their impact on system stability, the system’s ability to handle induced failures, and the growth and maturity of the organization’s chaos engineering practices. To address  \textbf{RQ3.3}, we categorized evaluation approaches into quantitative and qualitative methods based on themes derived from our systematic coding and thematic analysis (see Section~\ref{sec:thematic} and detailed mappings in the online appendix~\ref{sec:appendix}).

\paragraph{Quantitative Evaluation}

This involves systematically measuring numerical data to assess a system's performance under failure conditions. It focuses on metrics that can be measured, such as response time, error rates, and system availability~\cite{garg2013framework}. By comparing the behavior of the system during chaos experiments with its regular operation, engineers can identify weaknesses and areas for improvement based on objective data. We identified eight quantitative evaluation approaches, which include the following:

\begin{itemize}
    \item \textbf{Response Time Metrics} measure the total time it takes for a system to process a request, from when it is received to when the result is returned, encompassing both processing and response delivery~\cite{frank2021scenario,Abdul2024,frank2021interactive,Dua24,cotroneo2022thorfi,Green2023,Kostic2024,Mooney23,siwach2022evaluating,camacho2022chaos,sondhi2021chaos,NationalAustraliaBank20,al2024exploring,kesim2020identifying}. For example, a baseline response time of 18 milliseconds might increase to 27 milliseconds during chaos experiments such as 'KillNodes', where parts of the system are intentionally shut down to test disruptions, indicating slower request processing and completion~\cite{torkura2021continuous}.
    
    \item \textbf{Latency Metrics} measure the initial delay or waiting time between when a request is made and when the system begins processing that request~\cite{fogli2023chaos,Wickramasinghe23,Kostic2024,poltronieri2022chaos,Bremmers21,cotroneo2022thorfi,AzureReadiness23,Colyer19}. It does not include the total processing time, only the delay before the system starts responding~\cite{cotroneo2022thorfi,Durai22,Fawcett2020,Gianchandani2022,frank2021scenario,siwach2022evaluating,yu2021microrank}. For example, latency can increase from 44 to 70 milliseconds during chaos experiments, signaling that the system takes longer to react to requests~\cite{kassab2022c2b2}.
    
    \item \textbf{Resource Utilization Metrics} track the consumption of system resources, such as CPU and memory~\cite{AzureReadiness23,fogli2023chaos,torkura2020cloudstrike,simonsson2021observability,zhang20213milebeach,camacho2022chaos,pulcinelli2023conceptual,Wu21,Gianchandani2022,Roa2022,Kadikar2023,Green2023,zhang2019chaos}. For example, CPU usage may spike from 50\% to 80\% during chaos experiments, indicating an increased strain on the system due to failures ~\cite{simonsson2021observability}.
    
    \item \textbf{Error Rate Metrics} measure the frequency of errors encountered during system operation~\cite{Gill21,zhang20213milebeach,cotroneo2022thorfi,basiri2019automating,rivera2023using,Roa2022,Kostic2024,Vinisky2024,Colyer19,Durai22}. For instance, an increase in the error rate from 0.5\% to 5\% during chaos experiments suggests that the system is struggling to handle failures effectively~\cite{zhang2023chaos}.
    
    \item \textbf{Availability Metrics} refers to a system's ability to remain operational, measured by uptime and downtime~\cite{torkura2020cloudstrike,Gill21,camacho2022chaos,nikolaidis2021frisbee,ma2023phoenix}. Uptime indicates how long the system is fully functional, while downtime shows when the system is unavailable~\cite{Shah21,Durai22,Abdul2024,Gianchandani2022}. 
    
    \item \textbf{Throughput Metrics} measure the volume of data or transactions the system can handle in a given time~\cite{Kostic2024,Vinisky2024,al2024exploring,torkura2020cloudstrike}. For example, throughput might drop from 1,000 transactions per second to 600 during chaos experiments, leading to delays in processing user requests or data transfers~\cite{kassab2022c2b2,Abdul2024}.
    
    \item \textbf{Mean Time to Recovery (MTTR)} measures the average time required to restore the system after a failure~\cite{Durai22,Abdul2024,Starr22,Gianchandani2022}. A higher MTTR, indicating slower recovery, reveals areas that need faster recovery processes~\cite{Starr22,frank2021interactive,camacho2022chaos,Bhaskar22}. For example, if a cloud-based e-commerce platform experiences a server crash during a flash sale and MTTR increases from 2 to 10 minutes, it signals the need for more efficient recovery mechanisms to prevent revenue loss and customer frustration~\cite{siwach2022evaluating}.
    
    \item \textbf{Mean Time Between Failures (MTBF)} refers to the average time that a system operates without failure, reflecting the reliability of the system~\cite{Bhaskar22,camacho2022chaos}. A lower MTBF suggests frequent failures, highlighting the need for improved stability~\cite{Kadikar2023}. For example, if a healthcare system that manages patient records experiences failures every two hours, it points to the need for a more reliable system architecture to ensure uninterrupted access to patient data during critical procedures.  
    
\end{itemize}

\paragraph{Qualitative Evaluation}
This gathers non-numerical insights to assess system performance during failures, focusing on user feedback, team experiences, and business impacts. By understanding these aspects, engineers gain deeper insight into the broader effects of disruptions. This complements quantitative methods by adding meaning to numerical data. We identified four qualitative evaluation approaches, which include:

\begin{itemize}
    \item \textbf{User Experience Evaluation} involves gathering qualitative feedback on how system incidents impact end users~\cite{Gill21,ji2023perfce}. This may include conducting surveys, interviews, or usability tests to assess how failures affect users’ ability to interact with the system~\cite{camacho2022chaos,Kostic2024}. For example, users might report frustration or difficulty in completing tasks during a chaos experiment that causes slow page load times or feature outages~\cite{Fawcett2020,Sbiai2023,Gill2022}. This evaluation is crucial because it reveals how technical failures affect user satisfaction and retention, providing insight beyond response time or availability figures~\cite{Wu21,Le22,cotroneo2022thorfi}.
    \item \textbf{Business Impact Evaluation} assesses how chaos experiments affect key business metrics, including revenue, customer churn, and operational efficiency~\cite{Gremlin23,jernberg2020getting}. This involves understanding specific financial losses, customer complaints, and service-level breaches resulting from experiments\cite{NationalAustraliaBank20,Palani23}. For instance, a chaos experiment that causes downtime during a product launch could result in significant lost sales and a spike in customer service tickets~\cite{ji2023perfce,Starr22}. This evaluation is critical to linking technical failures with direct business consequences, enabling more targeted investment in stability measures where it matters most~\cite{Manna2021,Moon22}.
    \item \textbf{System Behavior and Observability} focuses on how effectively system components behave and are monitored during chaos experiments~\cite{Anwar19,yu2021microrank,meiklejohn2021service,Mooney23}. It involves reviewing logs/traces and monitoring gaps to identify areas where failures are not easily detected or diagnosed~\cite{ikeuchi2020framework,soldani2021automated,zhang2019chaos,Green2023}. For example, an experiment might expose missing alerts or inaccurate telemetry data for critical services, making it harder for engineers to respond~\cite{Colyer19,meiklejohn2021service,basiri2019automating}. This evaluation helps to ensure that the system's observability tools are robust enough to detect problems early and aid in faster diagnosis and recovery~\cite{dedousis2023enhancing,Fawcett2020}.
    \item \textbf{Team Feedback} captures insights from engineering teams on their ability to respond to failures during chaos experiments~\cite{Kostic2024,fogli2023chaos,torkura2020cloudstrike}. It evaluates the effectiveness of incident response, communication, and decision-making processes under system stress~\cite{zhang2019chaos,simonsson2021observability}. For example, feedback might highlight confusion in handling specific failure scenarios or delays in coordinating recovery efforts~\cite{Gremlin23, Krishnamurthy21,chen2022big}. This evaluation is essential to identify operational weaknesses in real-time incident management and to improve team workflows and readiness for actual outages~\cite{kesim2020identifying, Wu21, Fawcett2020}.
\end{itemize}

\section{Current State of Chaos Engineering (RQ4)}
\label{sec:Current_state}
This section addresses \textbf{RQ4} by exploring how chaos engineering has evolved in both academic and industry contexts. Specifically, it first presents the classification of research in chaos engineering, then examines the number of sources published per year and venue type \textbf{(RQ4.1)}, and finally reviews the venue rankings and key contributors in the field \textbf{(RQ4.2)}.
\subsection{Number of Sources by Research Type (RQ4.1)}
To better understand the current state of chaos engineering research (addressing \textbf{RQ4.1}), we classified the primary studies based on the research classification framework proposed by Wieringa et al.~\cite{wieringa2006requirements}. 
This framework includes the following categories:  
\begin{itemize}
    \item \textit{Evaluation research}: Investigates real-world problems through empirical methods such as case studies and surveys.
    \item \textit{Philosophical papers}: Propose new conceptual models, frameworks, or perspectives.
    \item \textit{Solution proposals}: Present novel techniques or tools, typically without full empirical validation.
    \item \textit{Validation research}: Rigorously assess proposed solutions in controlled environments (e.g., simulations, experiments).
    \item \textit{Opinion papers}: Express the author’s viewpoint on a topic, often arguing for or against certain practices.
    \item \textit{Experience reports}: Describe applied practices and lessons from real-world projects.
\end{itemize}
The first author did the classification, which was then reviewed by the second author, with all discrepancies resolved through discussion. The results of this classification, shown in Figure~\ref{fig:fig1}, indicate that solution proposal papers were the most frequent type of research between 2016 and 2024. In the academic literature, solution proposal papers were the most prevalent, with nine papers in 2021 and 2022, and slightly fewer (eight) in 2023. On the other hand, the grey literature showed a slight upward trend in experience reports, rising from six in 2022 to eight in 2023. This classification highlights the evolving distribution and focus of research activities, emphasizing the dominance of solution proposals in academic literature and the increasing role of practical experience reports in the grey literature.
\begin{figure}[!htbp]
  \centering
    \vspace{-10pt}
  \includegraphics[width=0.65\textwidth]{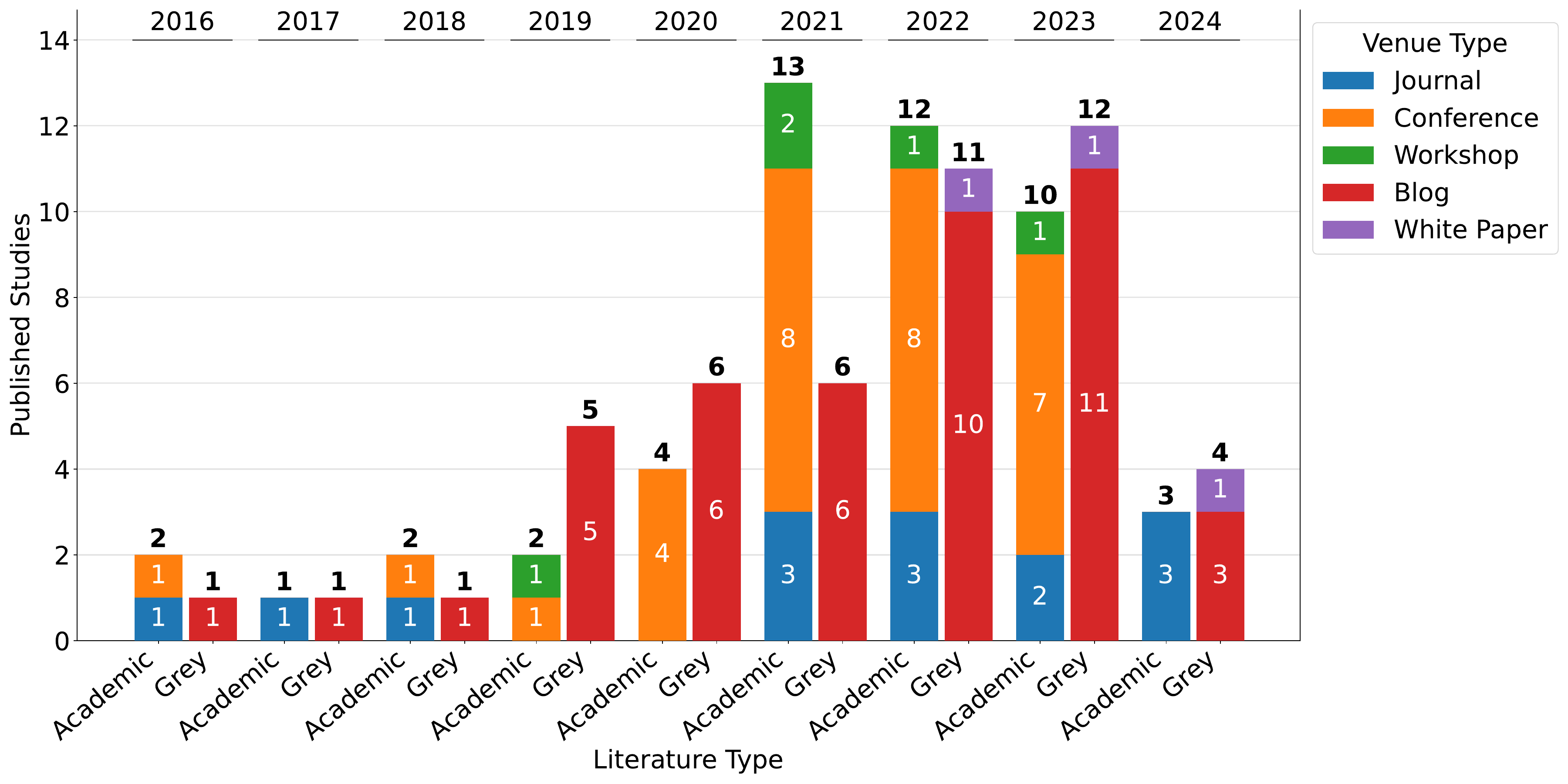}
    \vspace{-10pt}
  \caption{Distribution of Academic and Grey Literature Studies over Years and Venue Type.}
    \label{fig:Venueee}
  \Description{Distribution of academic(N = 49) and grey(N = 47) literature studies over years and venue type.}
\end{figure}
\subsection{Number of Sources by Year and Venue Type}
To further address \textbf{RQ4.1}, we analyzed the yearly distribution of academic and grey literature on chaos engineering from 2016 to early 2024 (see Figure~\ref{fig:Venueee}). Academic output was limited from 2016 to 2018 but grew significantly from 2019, peaking in 2021 and 2022 with 13 and 12 studies, respectively, mainly in the form of conference papers. Grey literature remained minimal until 2019, rising steadily to a peak of 12 sources in 2023, dominated by blog posts and a few white papers. A slight drop in 2024 is likely due to the incomplete year~\cite{werbinska2023maintenance}. Overall, these trends underscore the growing scholarly and industrial interest in chaos engineering. Further breakdowns of publication types, venues, and source classifications are available in our online appendix (Section~\ref{sec:appendix}).
\subsection{Publication trends and key Contributors in Academic and Grey Literature (RQ4.2)}
We analyzed publication trends in chaos engineering by examining venues based on their type and ranking and identifying key contributors within academic and grey literature. This analysis provides insights into the maturity of the domain~\cite{varela2021smart}.
\subsubsection{Academic Literature}
\begin{figure}[!htbp]
    \centering
    \begin{minipage}[t]{0.45\textwidth} 
        \centering
        \includegraphics[width=\textwidth]{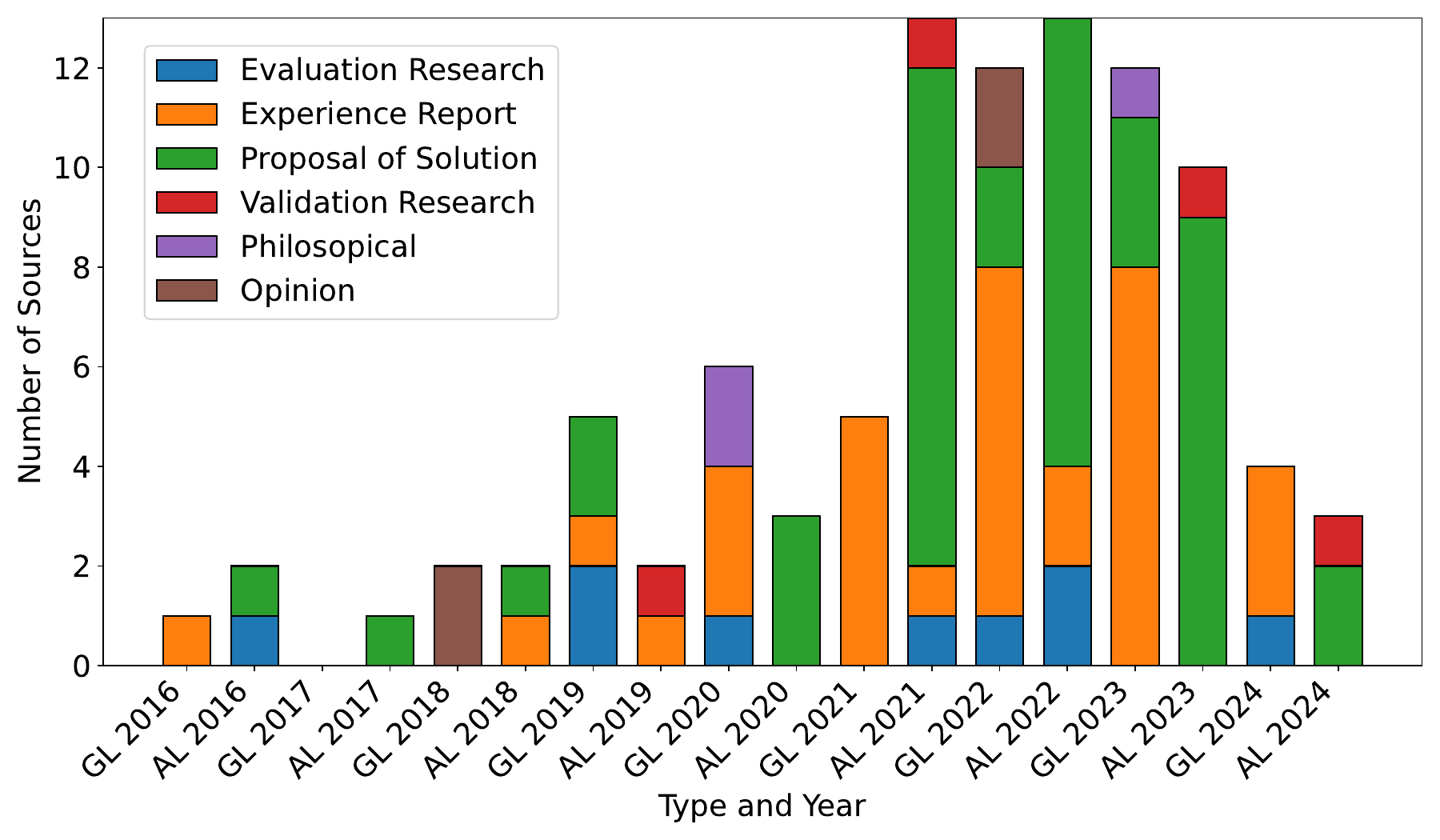}
        \Description{Research Type}
        \caption{Research Type of Academic (AL) and Grey (GL) Literature Studies.}
        \label{fig:fig1}
    \end{minipage}%
    \hspace{0.05\textwidth} 
    \begin{minipage}[t]{0.45\textwidth} 
        \centering
        \includegraphics[width=\textwidth]{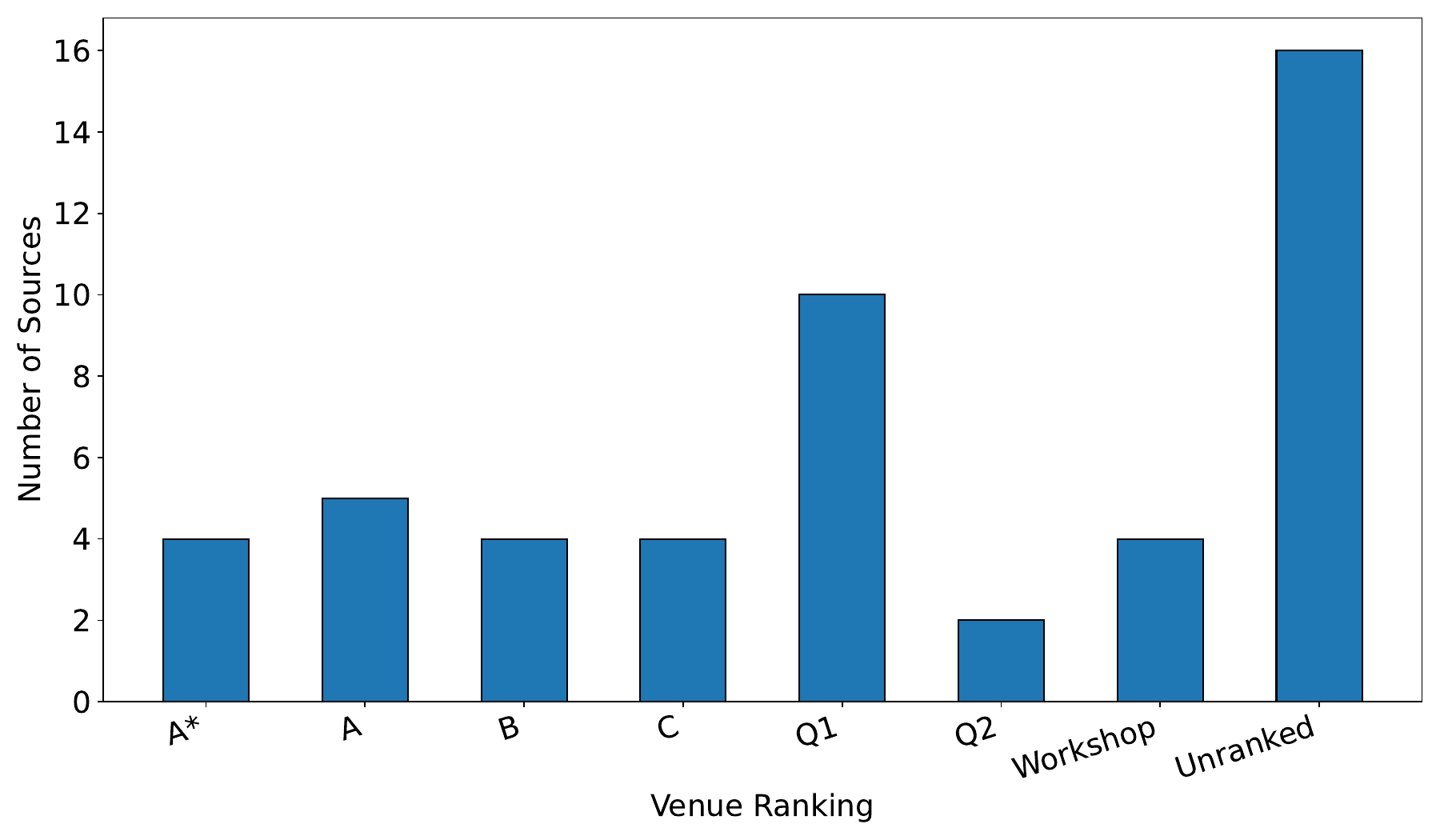}
        \Description{Venue Diagram}
         \caption{Venue Ranking of Academic Literature Studies.}
        \label{fig:fig2}
    \end{minipage}
\end{figure}

First, we analyzed publication venues based on their names, types (journal or conference), and rankings. The venues were categorized using the Computing Research and Education Association of Australasia (CORE) ranking for conferences (CORE 2023)\footnote{\url{https://portal.core.edu.au/conf-ranks/}} and the Scimago ranking for journals (Scimago 2023)\footnote{\url{https://www.scimagojr.com/journalrank.php}}. As shown in Figure~\ref{fig:fig2}, the largest share of studies appeared in unranked venues (12), though many were published in top-tier sources—particularly CORE A/A* conferences and Scimago Q1 journals, often from IEEE and ACM. This suggests increasing scholarly validation of chaos engineering research. Lower-ranked venues (B, C) and workshops contributed modestly, with few studies published in Q2 venues. To highlight influential contributions, we identified the top 10 academic studies based on citation counts from Scopus and Google Scholar (first author only). The leading study, authored by Netflix~\cite{basiri2016chaos}, has become foundational to the field, followed by diverse works from institutions in China, Germany, Sweden, and the U.S. that span topics such as end-to-end latency localization (e.g., MicroRank)~\cite{yu2021microrank}, chaos-based security testing~\cite{torkura2020cloudstrike}, and exception handling in JVM environments~\cite{zhang2019chaos}. These publications reflect a global effort toward improving system resiliency and reliability through chaos engineering. Full details are provided in our online appendix (Section~\ref{sec:appendix}).
\subsubsection{Grey Literature}
Unlike academic publications, grey literature lacks formal rankings and peer review. To evaluate its quality and impact, we utilized alternative metrics (e.g., likes, comments, shares) and, when social engagement data were limited, applied qualitative criteria such as venue reputation, author expertise, and content clarity (see Table~\ref{tab:quality_assessment}). The most engaged grey literature source was “Chaos Monkey Upgraded” by Netflix ~\cite{Hochstein2016}, which received 443 likes and 2 comments. Other widely read posts include Lyft’s framework introduction (348 likes)~\cite{Shah21}, NAB’s work on observability (156 likes)~\cite{Mondal20}, and technical contributions from Expedia~\cite{Katirtzis22} and the Chaos Toolkit community~\cite{Miles19}. These practitioner-authored works offer valuable, experience-based perspectives that complement academic research.
Further details, including top-ranked articles and alt-metric data, are available in the online appendix (Section~\ref{sec:appendix}).

\section{Discussion}
\label{sec:Discussion}
In this section, we summarize the key findings of our review and discuss their implications across \textbf{RQ1} to \textbf{RQ4}. We also outline open issues in chaos engineering research and discuss potential threats to the validity of our study.

\subsection{Implications of Findings}
The analysis highlights five key implications for researchers and practitioners in chaos engineering:
\begin{itemize}{
\item \textbf{Extending Reliability Testing through Chaos Engineering}: 
Chaos engineering should be understood not as a replacement for established reliability or software testing techniques, but as an extension that focuses on system behavior under failure and uncertainty. While traditional methods, such as unit tests, integration tests, or stress tests, validate expected behaviors in controlled environments, chaos engineering introduces carefully scoped disruptions to examine how systems respond and recover from unexpected faults. This approach contributes a complementary dimension to reliability assessment, particularly in distributed or production-like environments where emergent failures often go undetected. Rather than displacing existing practices, chaos engineering enhances them by validating resilience under real-world operational stress.
\item \textbf{Targeted Tool Selection (RQ3.1)}: 
This study introduces a taxonomy that helps practitioners select chaos engineering tools by considering factors such as execution environment, automation mode, and deployment stage (see Table \ref{tab:Tax}). The taxonomy is intended to support informed decision-making by aligning tool capabilities with the specific infrastructure and operational needs of an organization. In practice, a practitioner would begin by assessing the architectural characteristics of their system, such as whether it runs on Kubernetes, serverless platforms, or hybrid cloud, and then identify tools from the taxonomy that match these conditions and the intended stage of testing, from early experimentation to full production deployment. Applying the taxonomy effectively also requires careful consideration of how well a tool integrates with existing workflows and team expertise. Practitioners must ensure that selected tools support their resilience objectives without introducing unnecessary complexity or risk. For example, using a highly automated tool in an immature testing environment could result in poorly scoped experiments or overlooked failure modes. Likewise, choosing tools that lack compatibility with the target infrastructure may lead to inconclusive or misleading outcomes. By mapping tools to their most appropriate use cases, the taxonomy can help mitigate such risks and promote more efficient and targeted adoption of chaos engineering practices across diverse technical landscapes.
\item \textbf{Standardized Practices (RQ3.2)}: The findings highlight the importance of a standardized approach in chaos engineering to ensure consistent and practical application across diverse environments. Following best practices, which contribute to reliable and predictable resilience outcomes, such as faster recovery times and minimal service disruption~\cite{kumara2021s}, enables organizations to build resilience and systematically manage risks. These practices, such as starting with small, controlled experiments, setting clear objectives, and automating processes where possible, support organizations in achieving reliable outcomes crucial for managing complex systems (see Table~\ref{tab:best_practices_metrics}). Conversely, neglecting these practices, such as misaligned experiment design or insufficient safety mechanisms (see Table~\ref{tab:bad_practices_metrics}), can lead to inconsistent testing and increased system vulnerability. These findings underscore the need to balance adherence to best practices with avoiding common pitfalls, ensuring that chaos engineering aligns with broader operational goals to foster stability and reliability.
\item \textbf{Insights through Quantitative and Qualitative Fusion (RQ3.3)}: 
Another key implication of our findings is that combining quantitative and qualitative evaluation methods enhances the assessment and improvement of system stability in chaos engineering. Quantitative metrics provide objective data on performance under stress, including error rates, mean time to resolution, and availability. At the same time, qualitative insights, including user experience and team feedback, reveal the broader impacts of system failures. This combined approach enables organizations to identify technical vulnerabilities and understand their user and business impacts, allowing for targeted investments in both system stability and operational efficiency. This alignment of improvements with business goals and user expectations facilitates the development of solutions that effectively enhance system resilience, reduce failure impact, and improve user experience.
\item \textbf{Growing Alignment of Research Theory and Practice (RQ4)}:
Our review empirically demonstrates the growing alignment between academic research and industry practice in chaos engineering. By classifying 96 primary sources, including both peer-reviewed papers and grey literature, we observed that solution proposal papers dominated academic publications between 2016 and 2023, while grey literature during the same period showed a slight increase in experience reports (see Figure \ref{fig:fig1}). This dual trend reveals a convergence between theoretical contributions and practitioner-led experimentation, further reflected in the types of publication venues: top-ranked academic venues published novel frameworks and techniques, while widely cited blog posts detailed real-world implementation case studies. By capturing both perspectives, our study illustrates how academic research and industry practices are increasingly aligned, with researchers drawing insights from real-world applications and practitioners adopting concepts that resonate with emerging scholarly work to guide system resilience strategies. This evolving interplay signals an opportunity for deeper collaboration across academia and industry to accelerate innovation in chaos experiment design, such as advancements in automated fault orchestration, address implementation gaps, such as the lack of robust rollback or abort mechanisms in several research prototypes, and broaden the real-world adoption of chaos engineering.}
\end{itemize}

\subsection{Open Research Issues and Practical Challenges in Chaos Engineering}

Our review highlights several ongoing challenges that limit the adoption and scalability of chaos engineering. These include cultural resistance, skill gaps, and resource overhead. The following subsections outline three key themes based on current literature and practice.

\subsubsection{Organizational and Cultural Challenges}
The findings highlight that organizational resistance, risk aversion, and a blame culture can pose significant barriers to the adoption of chaos engineering~\cite{Krishnamurthy21, Bocetta19}. Although chaos engineering promotes resilience, many teams perceive fault injection, especially in production, as disruptive and risky~\cite{Sbiai2023, zhang2019chaos, dedousis2023enhancing}. This perception discourages widespread adoption, as teams are often reluctant to participate in experiments that could disrupt production or damage service reliability~\cite{Krishnamurthy21, ji2023perfce}. Adopting chaos engineering requires a mindset shift: from protecting systems at all costs to embracing failure as a tool for improvement~\cite{Bocetta19, Bremmers21}. However, this shift is rarely supported by leadership and is often undermined by siloed teams, poor communication, and inconsistent reliability practices~\cite{Katirtzis22, Dua24, Abdul2024}. Without cultural alignment and collaboration, chaos engineering remains confined to isolated efforts with limited organizational impact~\cite{Sbiai2023, Tivey19}. While these challenges are frequently noted in industry reports, academic research has yet to explore them systematically. There is a need for empirical studies that examine how factors like culture, team structure, and leadership influence chaos engineering adoption. Research could also contribute by developing and validating readiness frameworks or maturity models to guide organizations in scaling these practices effectively.

\subsubsection{Skill and Expertise Gaps}
Chaos Engineering requires specialized skills that many teams currently lack, creating a major barrier to adoption~\cite{torkura2020cloudstrike, Kostic2024}. First, effective implementation depends on understanding distributed systems, runtime behavior, and formal models like metric temporal logic, areas in which most practitioners have limited training~\cite{frank2023verifying, nikolaidis2021frisbee}. Next, managing experiment configurations, such as complex YAML files with interdependent settings, demands precision and familiarity with system internals~\cite{kassab2022c2b2}. Furthermore, many chaos engineering practices rely on intuition rather than standardized procedures, leading to inconsistent outcomes and reduced reliability~\cite{nikolaidis2021frisbee, Vinisky2024}. Adding to this challenge, some experiments require interaction with low-level components, such as kernel modules or orchestration layers, which increases the technical entry barrier~\cite{Katirtzis22, Mooney23}. Although these challenges are recognized in both academic and practitioner sources, there is limited empirical research on how teams acquire and apply these skills in practice. Future work should focus on identifying specific knowledge gaps, evaluating effective training strategies, and designing practical tools and reusable templates to support broader and more consistent adoption~\cite{Bairyev2023, cotroneo2022thorfi}.

\subsubsection{Resource and Operational Constraints}
Chaos engineering demands significant time, computing resources, expertise, and cost, which often limits routine adoption ~\cite{vu2022mission,poltronieri2022chaos,zhang20213milebeach}. Setting up environments, managing telemetry, and iterating experiments require sustained effort, especially for smaller or time-constrained teams~\cite{Gill21, zhang20213milebeach}. Deploying testbeds can take considerable time and need expert oversight~\cite{kassab2022c2b2}. These challenges scale with system size and are intensified by limited infrastructure, cloud expertise, and the financial overhead of tooling and operations~\cite{Katirtzis22, Durai22, Kostic2024}. As systems become increasingly complex, chaos engineering becomes more challenging to scale due to service heterogeneity and coordination overhead. Security and data concerns also arise. For example, experiments may expose vulnerabilities and generate sensitive telemetry. Managing this data poses storage, privacy, and analysis challenges. These factors—including time, complexity, and cost—often clash with delivery timelines, leading to chaos engineering being deprioritized despite its long-term value~\cite{Krivas2020, Kostic2024, Bairyev2023}. Although frequently reported in practice, there is little empirical research on how teams manage these constraints. Future work could investigate how organizations reduce the setup overhead of test environments or prioritize experiments to align with available resources. Such studies would help tailor chaos engineering to practical operational limits.


\subsection{Threats to Validity}

We considered potential threats to the external, construct, internal, and conclusion validity~\cite{wohlin2012experimentation} that may impact our study. 

\subsubsection{Threats to External Validity}

External validity refers to the extent to which our findings can be generalized beyond the specific context of our study~\cite{wohlin2012experimentation}. While this multivocal literature review (MLR) included both academic and grey sources, there is a possibility that some perspectives or practices in chaos engineering may not have been captured.
A particular threat arises from the risk of excluding relevant literature that uses alternative terminology, such as \textit{resilience testing} or \textit{fault injection}, rather than \textit{chaos engineering}. While we crafted inclusive search strings using multiple synonyms and related terms (see Table \ref{tab:search_results}), some relevant works may still have been omitted. This limitation is common in MLRs and is explicitly acknowledged. Future studies may improve coverage by expanding search terms to encompass other reliability testing practices.
Another limitation stems from the taxonomy itself, which has not yet undergone formal external validation. Although developed through thematic analysis and informed by structured internal review sessions, it remains untested in practical or industry settings. As such, its generalizability should be interpreted with caution until further validation studies are conducted.
Finally, our tool analysis focused on ten widely adopted chaos engineering tools due to feasibility constraints. While these tools represent current practitioner usage, broader tool coverage, including lesser-known or homegrown solutions, could yield deeper insights. We consider this an avenue for future work.

\subsubsection{Threats to Construct and Internal Validity}

Construct validity concerns whether the study accurately captures the concepts it aims to investigate, while internal validity refers to the soundness of the methods used and the control over potential biases~\cite{wohlin2012experimentation}.
To address these concerns, we conducted at least four structured feedback sessions during our analysis. Discussions from these sessions were qualitatively examined and used to fine-tune both our methodological choices and the relevance of our findings. We have also compiled an online appendix (see Section \ref{sec:appendix}) that contains all key artifacts, including the full list of sources, extracted codes, thematic groupings, and taxonomy classifications, to ensure transparency and reproducibility.
We also adopted triangulation by synthesizing both academic and grey literature sources, thereby reducing our reliance on a single perspective. Inter-rater reliability was formally assessed during the study selection phase, using Cohen’s Kappa, which indicated strong agreement for both academic and grey literature.
This ensured consistency in applying inclusion and exclusion criteria.
The first author conducted manual coding, which was independently reviewed by the second author. Disagreements were resolved collaboratively through discussion. This collaborative process was applied throughout all coding phases to ensure consistency and reduce interpretation bias.
Despite these precautions, some risk of observer bias remains inherent in qualitative research. However, we believe that our triangulation of sources, structured collaboration, and transparent artifact release significantly mitigate this threat.

\subsubsection{Threats to Conclusion Validity}

Threats to conclusion validity concern the degree to which the study’s conclusions are reasonably based on the data analyzed~\cite{wohlin2012experimentation}. To mitigate this, we applied thematic coding using \textit{ATLAS.ti} tool, and employed a structured analysis process to reduce observer and interpretation biases. Coding and classification were guided by a consensus-driven protocol between the first and second authors, ensuring consistency in the derivation of findings.
Both authors independently formulated the conclusions presented in this study and later validated them through collaborative discussion. These conclusions were cross-checked against the original source material to ensure they remained grounded in the data.
While our synthesis integrated both academic and grey literature, we acknowledge that our interpretations are limited to the scope of the selected articles. In particular, the absence of external expert validation for the taxonomy remains a limitation. Future research may benefit from follow-up evaluations, broader empirical testing, or practitioner feedback to confirm the applicability of our findings in real-world settings.
We also acknowledge that our results are limited to the quality of the grey literature and peer-reviewed publications, which we mitigate by conducting a quality assessment analysis.

\section{Conclusion and Future Work}
 \label{sec:Conclusion}

As modern distributed systems become increasingly complex, chaos engineering offers a proactive approach to ensure stability by introducing controlled failures that reveal system weaknesses, prevent outages, and support high performance under stress. This study bridges the gap between research insight and industry practices through a multivocal literature review (MLR) of 96 sources from academic and grey literature. To promote clarity across academia and industry, we proposed a unified definition of chaos engineering, reinforcing its role as a vital complement to traditional testing. We identified the main functionalities of a chaos engineering platform and compiled a set of activities that can guide systematic planning and conducting chaos experiments. In addition, we presented the key components of a chaos engineering platform and identified eleven quality requirements to guide the selection and evaluation of the platform. We also identified eight qualitative and four quantitative metrics essential for evaluating the impact of chaos experiments, ensuring that chaos engineering meaningfully enhances system robustness. Our review highlights three technical and four socio-technical challenges that chaos engineering seeks to resolve. To support organizations in structuring chaos engineering practices, we developed a taxonomy that organizes tools and techniques by environment, automation mode, automation strategy, and deployment stage while identifying the ten most commonly used tools within this framework. This categorization is valuable to practitioners and researchers when selecting tools that align with specific infrastructure needs and risk profiles.
In addition, we identified six best practices and five common pitfalls in the implementation of chaos engineering. These practices and potential pitfalls guide organizations in optimizing their chaos engineering initiatives to foster safer and more robust systems. Our findings also reveal that solution-oriented research in academia and practical experience in industry converge, underscoring the essential role of chaos engineering in strengthening modern distributed systems. Finally, we outlined several research and practical challenges that must be addressed to realize the vision of chaos engineering fully. We believe that these identified open issues can guide future research in the domain of chaos engineering.

In future work, we plan to use GitHub repository mining to identify various open-source projects that implement chaos engineering practices and utilize associated tools. This exploration will complement the ten tools analyzed in this literature review, offering broader insights into the prevalence and practical application of chaos engineering in diverse projects. Additionally, we aim to apply chaos engineering to enhance the robustness of AI-enabled systems by evaluating how these systems respond to disruptions in accuracy, adaptability, and consistency, ensuring they continue to deliver reliable performance even under challenging conditions.

\bibliographystyle{ACM-Reference-Format}

\bibliography{manuscript}

\appendix

\end{document}